\documentclass[apj]{emulateapj}
\usepackage[colorlinks,linkcolor={blue},citecolor={blue},urlcolor={red}]{hyperref}
\usepackage{epsfig,graphicx,natbib,amsmath,amsfonts,amssymb,epstopdf}%,xfrac}

\newcommand{\re}{\Reff}
\newcommand{\msolar}{M$_\odot$}
\newcommand{\mstar}{M$_\ast$}
\newcommand{\lgmstar}{$\log_{10}$(M$_\ast/$\msolar)}
\newcommand{\dindex}{D$_n$(4000)}
\newcommand{\nuvr}{NUV$-r$}
\newcommand{\hd}{H$\delta$}
\newcommand{\hda}{\hd$_A$}
\newcommand{\ewhda}{EW(\hda)}
\newcommand{\ha}{H$\alpha$}
\newcommand{\hae}{\ha}
\newcommand{\ewhae}{EW(\hae)}
\newcommand{\lgewhae}{$\log_{10}$\ewhae}

\newcommand{\Reff}{{$R_{\rm e}$}}

\newcommand{\myemail}{\email{ecwang16@ustc.edu.cn (EW); cli2015@tsinghua.edu.cn (CL)}}

\shorttitle{Gradients in recent star formation histories}
\shortauthors{Wang et al.}

\graphicspath{{fig/}{fig/com_par/}{fig/example/}{fig/D4000/}{fig/diagram/}{fig/dependence/}{fig/discussion/}{fig/profile/}{fig/sample/}{fig/com_slope/}{fig/resmass/}{fig/check/}{fig/fq_intro/}}
\begin{document}

\title{SDSS-IV MaNGA: Star formation cessation in low-redshift galaxies I. Dependence on stellar mass and structural properties}
\author{
Enci Wang\altaffilmark{1,3,4},
Cheng Li\altaffilmark{2,3}
Ting Xiao\altaffilmark{3,5}, 
Lin Lin\altaffilmark{3},
Matthew Bershady\altaffilmark{6},
David R. Law\altaffilmark{7},
Michael Merrifield\altaffilmark{8}, \\
Sebastian F. Sanchez\altaffilmark{9}, 
Rogemar A. Riffel\altaffilmark{10,11},
Rogerio Riffel\altaffilmark{12,11},
and Renbin Yan\altaffilmark{13}
} 
\myemail

\altaffiltext{1}{CAS Key Laboratory for Research in Galaxies and Cosmology, Department of Astronomy, University of Science and Technology of China, Hefei 230026, China; ecwang16@ustc.edu.cn}
\altaffiltext{2}{Tsinghua Center for Astrophysics and Physics Department, Tsinghua University, Beijing 100084, China; cli2015@tsinghua.edu.cn}
\altaffiltext{3}{Shanghai Astronomical  Observatory,  Nandan Road  80, Shanghai  200030, China}
\altaffiltext{4}{School of Astronomy and Space Science, University of Science and Technology of China, Hefei 230026, China}
\altaffiltext{5}{Department of Physics, Zhejiang University, Hangzhou 310027, China}
\altaffiltext{6}{Department of Astronomy, University of Wisconsin-Madison, 475 N. Charter Street, Madison, WI, 53706, USA}
\altaffiltext{7}{Space Telescope Science Institute, 3700 San Martin Drive, Baltimore, MD 21218, USA}
\altaffiltext{8}{School of Physics \& Astronomy, University of Nottingham, Nottingham NG7 2RD, UK}
\altaffiltext{9}{Instituto de Astronom\'ia, Universidad Nacional Aut\'onoma de  M\'exico, A.~P. 70-264, C.P. 04510, M\'exico, D.F., Mexico }
\altaffiltext{10}{Departamento de F\'\i sica, Centro de Ci\^encias Naturais e Exatas, Universidade Federal de Santa Maria, 97105-900, Santa Maria, RS, Brazil}
\altaffiltext{11}{Laborat\'orio Interinstitucional de e-Astronomia - LIneA, Rua Gal. Jos\'e Cristino 77, Rio de Janeiro, RJ - 20921-400, Brazil}
\altaffiltext{12}{Departamento de Astronomia, Instituto de F\'\i sica, Universidade Federal do Rio Grande do Sul, CP 15051, 91501-970, Porto Alegre, RS, Brazil}
\altaffiltext{13}{Department of Physics and Astronomy, University of Kentucky, 505 Rose Street, Lexington, KY 40506, USA}

\begin{abstract} 
We investigate radial gradients in the recent star formation history (SFH) of low-redshift galaxies using a large sample of 1917 galaxies with $0.01<z<0.14$ and integral-field spectroscopy from the ongoing Mapping Nearby Galaxies at Apache Point Observatory (MaNGA) survey. For each galaxy, we obtain two-dimensional maps and radial profiles for three spectroscopically-measured parameters that are sensitive to the recent SFH: \dindex\ (the 4000\AA\ break), \ewhda\ (equivalent width of the \hd\ absorption line), and \ewhae\ (equivalent width of the \ha\ emission line). We find the majority of the spaxels in these galaxies are consistent with models with continuously declining star formation rate, indicating that starbursts occur rarely in local galaxies with regular morphologies. We classify the galaxies into three classes: fully star-forming (SF), partly quenched (PQ) and totally quenched (TQ), according to the fraction of quenched area within 1.5 times the effective radius, $f_Q(1.5R_e)$. We find that galaxies less massive than a stellar mass of $10^{10}$\msolar\ present at most weak radial gradients in all the diagnostic parameters. In contrast, massive galaxies with stellar mass above $10^{10}$\msolar\ present significant gradients in all the three diagnostic parameters if classified as SF or PQ, but show weak gradients in \dindex\ and \ewhda\ and no gradients in \ewhae\ if in the TQ class. This implies the existence of a critical stellar mass, at $\sim10^{10}$\msolar, above which the star formation in a galaxy gets shutdown from the inside out. Galaxies tend to evolve synchronously from inner to outer regions before their mass reaches the critical value. We have further divided the sample at fixed mass by both bulge-to-total luminosity ratio and morphological type, finding that our conclusions hold regardless of these factors: it appears that the presence of a central dense object like a bulge is not a driving parameter, but rather a byproduct of the star formation cessation process.

\end{abstract}


\section{Introduction}
\label{sec:introduction}

%Galaxies in the local Universe can be well divided into two major populations according to their UV/optical colors \citep[e.g.][]{Strateva-01, Blanton-03, Baldry-04} or specific star formation rates \citep[e.g.][]{Kauffmann-03b, Brinchmann-04}: a population of actively star-forming galaxies located in the ``blue cloud'' of the color-mass diagram, and a population of quiescent galaxies located in the ``red sequence'' of the same diagram. The galaxy bimodality was originally discovered from low-z galaxies in the Sloan Digital Sky Survey \citep[SDSS;][]{York-00}, and has been found to persist out to redshifts of at least 2.5 based on narrow-field deep imaging surveys \citep[e.g.][]{Bell-04, Bundy-06, Cirasuolo-07, Faber-07, Martin-07, Cooper-08, Cowie-Barger-08, Brammer-09, Williams-09, Brammer-11,  Muzzin-12, Huang-13, Tomczak-14}. The fraction of red-sequence galaxies in the universe has gradually, but significantly increased since those redshifts, implying that blue-cloud galaxies at high-z have largely evolved to the present-day red-sequence galaxies. Therefore, the cessation of star formation is an important process which has been driving the evolution of the galaxy population since $z=2.5$ or even higher redshifts. 

The fraction of red, quiescent galaxies in the universe has gradually, but significantly increased since $z\sim1$ \citep[e.g.][]{Bell-04, Bundy-06, Faber-07}, implying that the population of blue, star-forming  galaxies at high redshift have largely evolved to the present-day red population. The cessation of star formation is thus a driving  process for galaxy evolution over the past $\sim8$ Gyr. A complete picture of the way in which star formation gets shut down remains elusive, however, although extensive studies have been carried out in the past decades, both observational and theoretical. These studies have proposed and examined a variety of physical processes, finding both external environment and the internal structural properties to play important roles in star formation cessation in galaxies. 

%For instance, major mergers of gas-rich disks of comparable masses have long been proposed as an efficient channel of making massive quenched galaxies with elliptical morphology, which is also predicted to trigger powerful ``quasar-mode'' AGN feedback \citep[e.g.][]{DiMatteo-Springel-Hernquist-05, Hopkins-06}, commonly adopted as one of the quenching processes in semi-analytic models \citep[e.g.][]{Croton-06}. 

For central galaxies located at the center of massive dark halos with halo mass exceeding a critical mass of $\sim10^{12}$\msolar, shock heating may effectively suppress the gas cooling efficiency, thus preventing further star formation in massive central galaxies \citep[e.g.][]{Rees-Ostriker-77, Silk-77, Blumenthal-84, Birnboim-Dekel-03, Keres-05, Dekel-Birnboim-06, Cattaneo-06}. In this case, accretion of the hot halo gas onto the central super-massive black hole in central galaxies is expected to trigger ``radio-mode'' active galactic nucleus (AGN) feedback, commonly adopted as one of the quenching\footnote{We note that in this paper ``quenching'' doesn't specifically mean a rapid shutdown of star formation. In most cases we use ``cessation'' instead of ``quenching'', thus including both rapid shutdown and slow diminution over long timescales.} processes in semi-analytic models \citep[e.g.][]{Croton-06}. For centrals hosted by less massive halos, processes internal to the systems are more important. Observational studies have clearly established that a massive dense structure at galactic center (such as, a prominent bulge) is presented in the majority of quenched central galaxies, at both low and high redshifts \citep{Bell-08, Bundy-10, Masters-10a, Bell-12, Cheung-12, Fang-13, Bluck-14, Barro-17}. This finding was initially interpreted as observational evidence for AGN feedback quenching star formation in central galaxies \citep[e.g.][]{Bell-08}, and later on also thought of as a natural consequence of the ``morphology quenching'' picture in which a massive central object is able to stabilize the rotating disc against star formation \citep[e.g.][]{Martig-09}. 

In contrast, star formation cessation in satellite galaxies appears to be primarily driven by external processes occurring within or around the host groups. These processes include gas stripping by tidal interactions \citep{Toomre-Toomre-72, Moore-96} and ram pressure \citep{Gunn-Gott-72, Abadi-Moore-Bower-99, Vogt-04, Chung-09, Merluzzi-13}, galaxy mergers \citep[e.g.][]{Lin-14}, and more gentle processes such as ``starvation'' or ``strangulation'' \citep[e.g.][]{Balogh-Navarro-Morris-00, Weinmann-09, Peng-Maiolino-Cochrane-15}. The importance and strength of these environmental effects depend on a variety of factors: dark matter mass of the host halos, cluster-centric radius of galaxy locations, galaxy stellar mass, and particularly the structural parameters of galaxies (e.g. surface stellar mass density; \citealt{Li-12, Zhang-13}).

Therefore, in order to have a full understanding of star formation cessation in galaxies, one would need spatially resolved spectroscopy to measure the stellar and gaseous components for a large sample of galaxies covering wide ranges in global properties (e.g. stellar mass, color, star formation rate, morphology, nuclear activity, etc.), and probing environmental conditions over a wide range of spatial scales (tidal interactions/mergers, central/satellite classification, local density, dark halo mass, halo assembly history, large-scale structure configuration, etc.). Resolved spectroscopy has become available only recently thanks to the integral field unit (IFU) surveys accomplished in the past decade: SAURON \citep{Bacon-01, deZeeuw-02}, DiskMass \citep{Bershady-10}, ATLAS$^{\rm 3D}$ \citep{Cappellari-11}, CALIFA \citep{Sanchez-12}, VENGA\citep{Blanc-13}, SLUGGS\citep{Brodie-14} and MASSIVE\citep{Ma-14}. These IFU surveys have typically targeted tens or hundreds of local galaxies with different morphological types, enabling detailed studies of the spatially resolved stellar populations and kinematics of galaxies. Meanwhile, the KMOS$^{\rm 3D}$ survey is extending the IFU efforts to higher redshifts, observing 600 galaxies at $0.7<z<2.7$ using KMOS at the Very Large Telescope \citep{Wisnioski-15}. 

These IFU surveys have enabled two-dimensional maps of star formation and mass assembly histories to be studied with high accuracy for galaxies in the local universe. For instance, the CALIFA survey has spatially resolved light-weighted stellar age, star formation rate, oxygen abundance and stellar metallicity, and the history of stellar mass assembly for a few hundred galaxies with different morphologies and masses, revealing an inside-out growth picture for massive galaxies and a transition to outside-in growth at \mstar$\sim10^{10}$\msolar\  \citep{Perez-13, GonzalezDelgado-14, Sanchez-14, Sanchez-Blazquez-14, GonzalezDelgado-15, GonzalezDelgado-16, Catalan-Torrecilla-17}. Specifically, for massive galaxies, the intensity of SFR is found to decline as one goes from the galactic center outward, suggesting the cessation of star formation happens from inside out. This inside-out picture of star formation cessation is broadly consistent with previous studies, which have been based on either multi-wavelength broad-band photometry \citep[e.g.][]{deJong-96, Abraham-99, Bell-deJong-00, Munoz-Mateos-07, Zibetti-Charlot-Rix-09, Roche-Bernardi-Hyde-10, Suh-10, Tortora-10, Gonzalez-Perez-Castander-Kauffmann-11, Tortora-11, Szomoru-12, Szomoru-13, Wuyts-12, Wuyts-13, Kauffmann-15}, or long-slit spectroscopy \citep[e.g.][]{Moran-12, Huang-13}.

More recent IFU surveys have started to construct large statistical samples consisting of more than a thousand galaxies at low-$z$. The SAMI survey \citep{Croom-12} is observing 3400 galaxies using a multiplexed fiber-IFU instrument at the Australian Astronomical Observatory. The Mapping Nearby Galaxies at Apache Point Observatory \citep[MaNGA;][]{Bundy-15} project is making another big step forward by surveying 10,000 galaxies selected from the SDSS spectroscopic sample. The MaNGA sample spans a wide dynamic range in stellar mass, color, star formation rate and environment. More importantly, MaNGA has effective \'{e}tendue in the near UV (see \citealt{Bundy-15}), allowing precise measurements of spectral indices such as the 4000\AA\ break [\dindex] and the equivalent width of \hd\ absorption line [\hda], indicative of young stellar populations formed in the past $0.1-1$ Gyr \citep{Kauffmann-03a}. These stellar indices, together with the H$\alpha$ emission line equivalent width (\ewhae),   provide powerful diagnostics of the recent star formation history of given galactic region.  MaNGA thus has unique statistical power in studying star formation cessation in galaxies, and a comprehensive picture of the star formation cessation process may be built up by measuring gradients in these diagnostics for the large sample of galaxies from MaNGA.

\cite{Li-15} (hereafter Paper I) presented a demonstration of MaNGA's unique potential in studies of star formation cessation, analyzing the radial profiles of \dindex, \ewhda\ and \ewhae\ for 12 galaxies observed in the MaNGA prototype run (P-MaNGA; see \citealt{Bundy-15}). The radial variation in these diagnostics is found to depend strongly on the star formation status of the galactic center: a centrally star-forming galaxy generally shows no or very weak radial variation across the whole galaxy, while centrally quiescent galaxies present significant gradients in all the diagnostics, supporting the ``inside-out'' picture in which the cessation of star formation propagates from the center of a galaxy outward. In a parallel paper, \cite{Wilkinson-15} derived gradients in stellar age and metallicity by performing full spectral fitting to the P-MaNGA data, finding similar, negative gradients in stellar age in centrally quiescent galaxies. Using a representative sample of 721 galaxies from the MaNGA internal data release, MaNGA Product Launches 4 \citep[MPL4, equivalent to SDSS/DR13;][]{SDSS-DR13}, \cite{Goddard-17} extended the P-MaNGA analysis of \cite{Wilkinson-15} and found that the gradients in stellar age tend to be shallow for both early-type and late-type galaxies, with on average a positive mass-weighted age gradient ($\sim$0.09 dex/$R_e$) for early-type galaxies and a negative light-weighted age gradient ($\sim-$0.11 dex/$R_e$) for late-type galaxies.

This paper is the first of a series of papers in which we will extensively study the star formation cessation process in low-redshift galaxies as observed in MaNGA. In this paper, we extend the analysis of Paper I by using a much larger sample of 1917 galaxies from the MaNGA Product Launches 5 \citep[MPL5, equivalent to the SDSS/DR14;][]{SDSS-DR14}. Following Paper I, we use \dindex, \ewhda\ and \ewhae\ to trace the star formation history over the last 1-2 Gyr, for each spaxel in the MaNGA datacubes, thus obtaining both two-dimensional maps and one-dimensional radial profiles of the recent SFH diagnostics for the galaxies. When compared to Paper I, both the sample size and the diagnostic parameter measurements are largely improved, enabling us to measure the gradients in recent SFH over a wide dynamic range in galaxy mass, color and structural properties. Based on these measurements, we propose a new quantity, $f_Q$, which is the fraction of quenched area within 1.5$R_e$, to quantify the global status of the star formation in a galaxy. Accordingly, we classify our galaxies into three subsets at different stages in terms of global star formation: fully star-forming (SF), partly quenched (PQ) and totally quenched (TQ). We compare this new classifier with both global colors and \dindex\ at galactic center measured from the SDSS 3$^{\prime\prime}$-fiber spectroscopy, which have been widely adopted for the same purpose in previous studies \citep[e.g.][]{Kauffmann-03a, Kauffmann-07, Cheung-12, Pan-14, Guo-16}. We analyze the gradients in recent SFH for the three types of galaxies, particularly examining the dependence on stellar mass and structural parameters such as bulge-to-disk ratio. As we will show, these analyses provide interesting implications for the star formation cessation process which is found to depend strongly on the stellar mass of the galaxy. In following papers of the series, we will further investigate the roles of galactic bars,  galaxy-galaxy interactions/mergers and large-scale environment in regulating the star formation cessation of galaxies. We will also study the cold gas content of these galaxies by performing followup observations of the atomic and molecular gas emission. 
Throughout this paper, we assume a flat cold dark matter cosmology model with $\Omega_m=0.27$, $\Omega_\Lambda=0.73$ and $h=0.7$ when computing distance-dependent parameters.

\section{Data}
\label{sec:data}

\subsection{MaNGA overview}
\label{sec:data_sample}

As one of the three major experiments of the fourth generation of the Sloan Digital Sky Survey \citep[SDSS-IV;][]{Blanton-17}, MaNGA is obtaining integral-field spectroscopy for 10,000 nearby galaxies with redshifts in the range $0.01<z<0.15$ \citep{Bundy-15}. The MaNGA instrument utilizes the two dual-channel BOSS spectrographs at the 2.5m Sloan Telescope \citep{Gunn-06, Smee-13}, simultaneously covering wavelengths over 3600-10300 \AA\ at R$\sim$2000, and reaching a target $r$-band  $S/N=4-8$ (\AA$^{-1}$ per 2$^{\prime\prime}$-fiber) at 1-2$R_e$, with a typical integration time of 3 hr. The MaNGA IFU includes 29 fiber bundles for the Sloan 3$^\circ$-diameter field of view, including 12 seven-fiber ``mini-bundles'' dedicated for flux calibration and 17 science bundles with five different sizes ranging from 19 to 127 fibers, corresponding to on-sky diameters between 12$^{\prime\prime}$ and 32$^{\prime\prime}$. Details of the MaNGA instrument design, testing and assembly can be found in \cite{Drory-15}.

MaNGA raw data are reduced with the Data Reduction Pipeline developed by \citet{Law-15, Law-16}. The reduced data for each galaxy is provided in the form of a datacube with a spaxel size of 0\farcs5, which was chosen as a suitable compromise between galaxy spatial information and data volume. The typical effective spatial resolution of the reconstructed datacubes can be described by a Gaussian with a FWHM$\sim$2\farcs5. Flux calibration has mainly corrected for the flux loss due to atmospheric absorption and the instrument response. With the help of the ``mini-bundles'', the absolute flux calibration is better than 5\% for more than 89\% of MaNGA's wavelength range. Flux calibration, survey strategy and data quality tests are described in detail in \cite{Yan-16a, Yan-16b}. 

\cite{Wake-17} present the MaNGA sample design and optimization, which were developed so as to simultaneously optimize the MaNGA IFU size distribution, the IFU allocation strategy, and the target density. Three samples are defined. The Primary and Secondary samples are selected to have a flat distribution of the $i$-band absolute magnitude, $M_i$,  with the assigned IFU bundles reaching to 1.5 and 2.5 effective radii, respectively. Supplemental to the Primary sample, the Color-Enhanced sample increases the fraction of rare populations in the color-magnitude diagram, such as high-mass blue galaxies, low-mass red galaxies and ``green-valley'' galaxies, by extending the redshift limits in the appropriate color bins. 

An overview of the MaNGA instrument, survey design, key science goals and the prototype observations (P-MaNGA) are presented in \cite{Bundy-15}. The first MaNGA data release accompanied the SDSS Data Release 13 \citep{SDSS-DR13}, including reduced MaNGA datacubes for 1390 galaxies observed between July 2014 and July 2015. The SDSS Data Release 14 \citep{SDSS-DR14} has come out more recently, and provides MaNGA datacubes for a total of 2812 galaxies observed by July 2016. Two value-added catalogues (VACs) based on MaNGA data are released together with the SDSS/DR14, which are contributed by the scientists within the MaNGA team and contain spatially resolved and integrated stellar population properties derived by performing full spectral fitting of the MaNGA datacubes \citep{Sanchez-16a, Goddard-17}.

\subsection{SDSS perspective of the MPL5 sample}
\label{sec:sdss_view}

\begin{figure*}
  \begin{center}
    \epsfig{figure=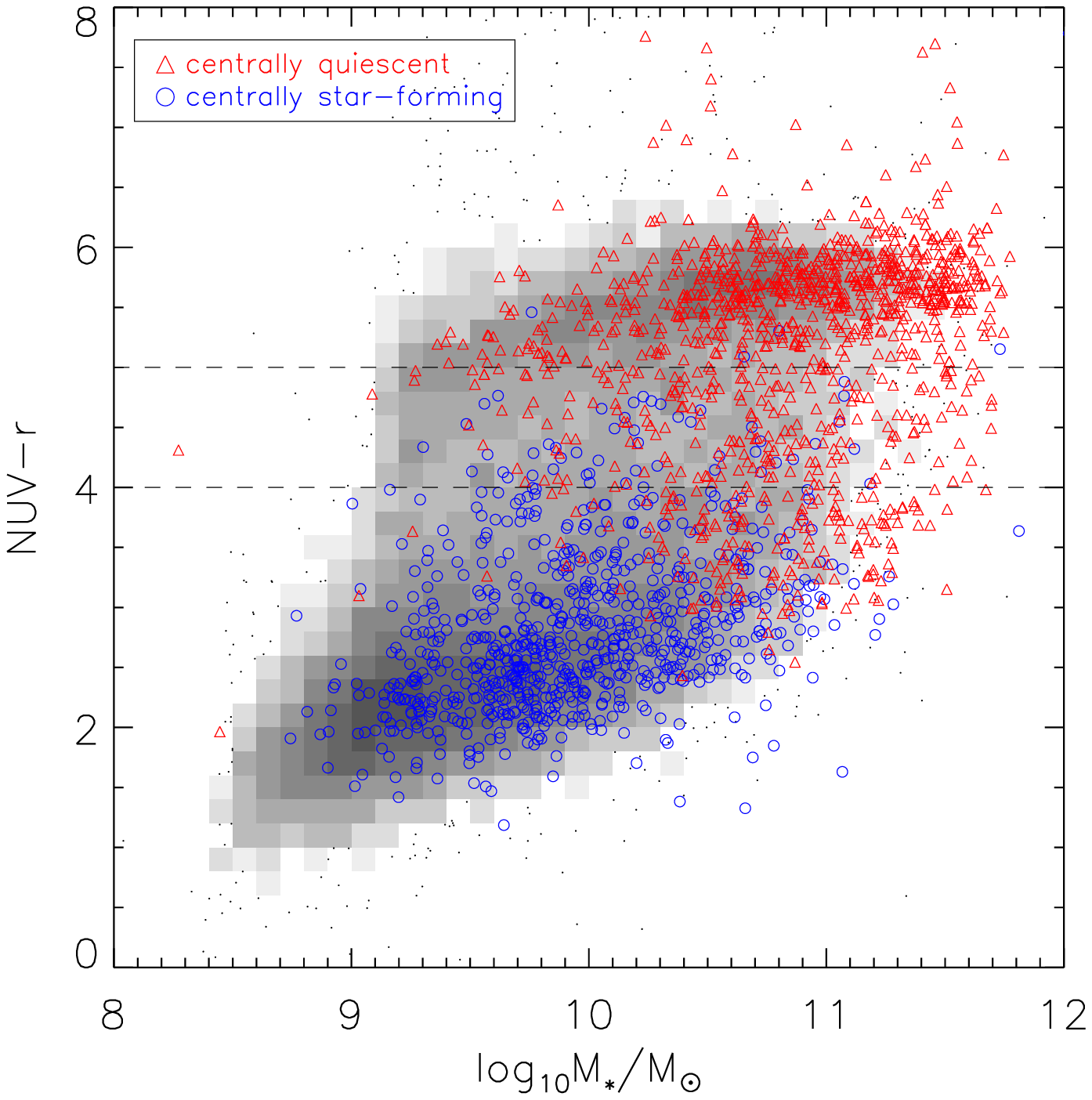,clip=true,width=0.4\textwidth}
    \epsfig{figure=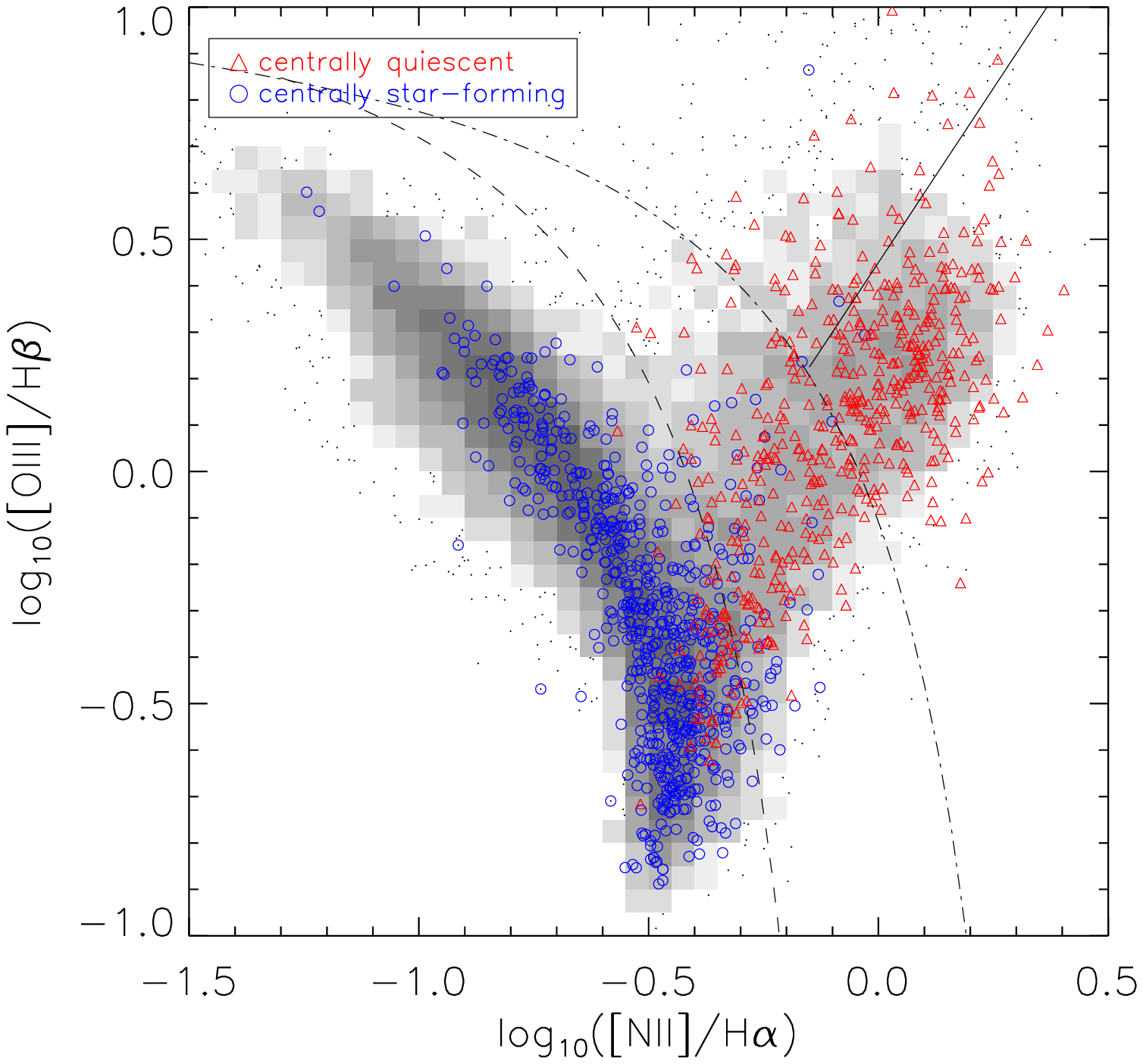,clip=true,width=0.4\textwidth}
  \end{center}
  \caption{Centrally quiescent (red triangles) and centrally star-forming 
   (blue circles) galaxies are plotted on the stellar mass vs. \nuvr\ plane (left panel), as well as the diagram of \citep{Baldwin-Phillips-Terlevich-81} (right panel). For comparison, a volume-limited sample selected from SDSS are shown in the background in both panels. The two horizontal lines in the left panel indicate \nuvr$=4$ and 5. In the right panel, the dashed and dashed-dotted curves are adopted from \citet{Kauffmann-03a} and \citet{Kewley-06} to separate AGN from star forming galaxies.}
  \label{fig:sample_properties}
\end{figure*}

\begin{figure*}
  \begin{center}
    \epsfig{figure=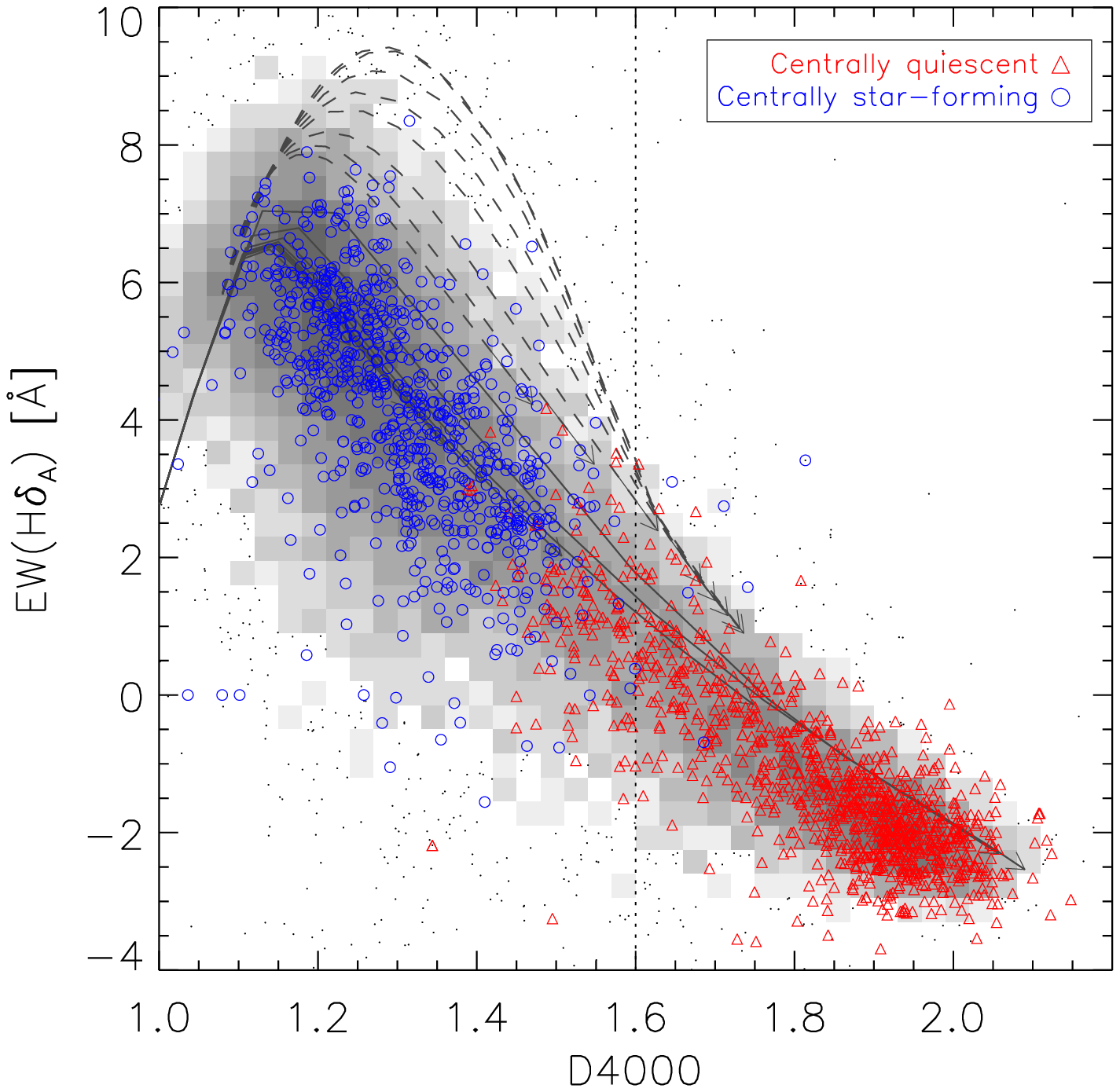,clip=true,width=0.33\textwidth}
    \epsfig{figure=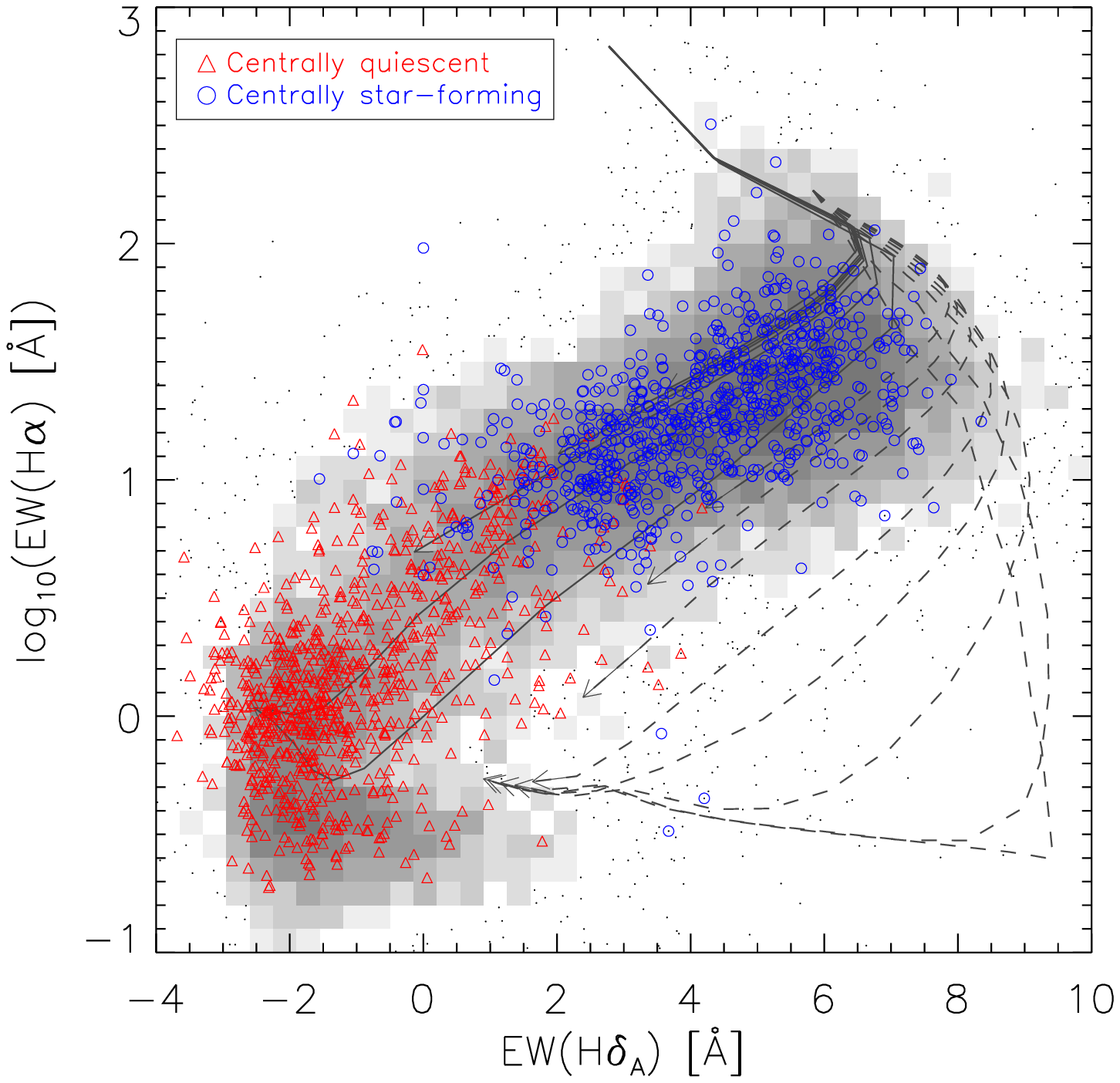,clip=true,width=0.33\textwidth}
    \epsfig{figure=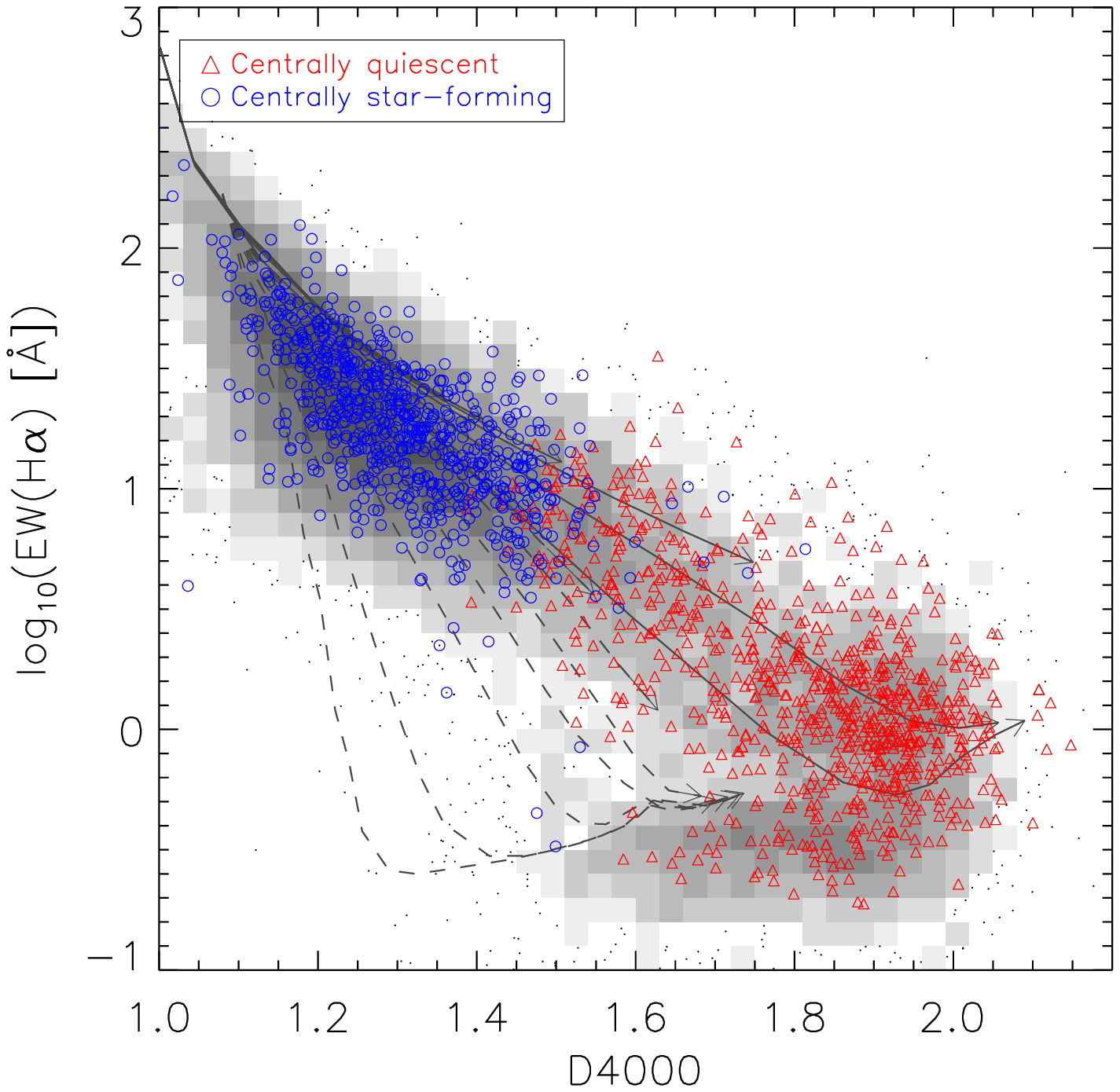,clip=true,width=0.33\textwidth}
  \end{center}
  \caption{Galaxies classified into centrally quiescent (red triangles) and
centrally star-forming (blue circles) are shown on planes of \dindex\ vs. \ewhda\
 (left   panel),    \ewhda\  vs. \ewhae\  (middle panel) and \dindex\ vs. \ewhae\ (right panel). The solid and dashed lines are predictions by solar metallicity models of \cite{Bruzual-Charlot-03} with continuously declining star formation 
 (solid) or instantaneous starbursts (dashed). For comparison, the distribution of a volume-limited sample selected from SDSS is shown as grayscale background in both panels.}
  \label{fig:sample_diagram}
\end{figure*}

In this work we make use of the MPL5 (the version-2-0-1 of MaNGA Product Launches) data, which is equivalent to the MaNGA sample released with SDSS/DR14, including a total of 2778 galaxies. We restrict ourselves to galaxies with regular shapes, thus excluding those with irregular/disturbed morphologies; (13.2\% of the galaxies are excluded for this reason), and will be studied in a separate paper of the series (T. Xiao et al. in preparation). The morphological type of each galaxy is identified by visually examining the $r$-band image from SDSS. We classify the sample galaxies into three broad classes: disk-like, spheroid-like and irregular. We have compared our classifications with those from the Galaxy Zoo project \citep{Willett-13}. Among the irregular/disturbed galaxies in our case, only 13.4\% are classified as regular morphology with the voted odd fraction $<$0.3 in Galaxy Zoo, indicating that the two classifications are broadly consistent. We find almost all the spheroid-like galaxies (97.8\%) in our classification are smooth-like in Galaxy Zoo. However, our spheroid-like classification is much stricter than the smooth-like classification in Galaxy Zoo project, since 38.2\% of the smooth-like galaxies in Galaxy Zoo are classified to be disk-like according to our classification. We have visually examined the galaxies with discrepant classifications and verified that they do have spiral-like features in their image. Furthermore, we exclude 475 galaxies from the Color-Enhanced sample and keep only those from the Primary and Secondary samples, considering that the $1/V_{max}$ weighting scheme is applicable only to these samples \citep[see][]{Wake-17}. Finally we excluded 19 galaxies with apparent problems in data reduction or spectral fitting. These restrictions yield a sample of 1917 galaxies in total. We take this as our full sample, and all the analyses below will be based on this sample.

In what follows, we will correct the sample selection effects with the $1/V_{max}$ weighting method following the prescription in \cite{Wake-17}. For a given galaxy in the Primary or Secondary sample, with an SDSS $i$-band magnitude of $M_i$, \cite{Wake-17}  have calculated the minimum redshift $Z_{min}$ and the maximum redshift $Z_{max}$, over which the galaxy can be included in the MaNGA target sample. This corresponds to a maximum volume, $V_{max}$. The MaNGA sample selection effects are then corrected by weighting each galaxy by $1/V_{\rm max}$ when performing statistical analyses. The weights are multiplied by a factor 1/0.769 for galaxies from the Secondary sample, which is randomly selected from the target sample at a rate of 76.9\% for real observations.  We note that we have ignored a slight selection bias in angular size caused by the fact that for multiple reasons the MaNGA IFU size distribution does not exactly match the targeted distribution (see \S6.2 in \citealt{Wake-17} for details), which is expected to have no significant effect on our analyses. Our sample covers a redshift range of 0.01$<$z$<$0.14, corresponding to a cosmic time interval of 1.6 Gyr. We assume the cosmic evolution of our measurements is negligible. In fact, we have repeated the main analyses to be presented in the rest of the paper, but limiting the sample to a narrower redshift range of z$<$0.06, and found the results remain essentially unchanged. 

Before analyzing the MaNGA datacubes, we have examined the global properties of our sample using the SDSS single-fiber spectroscopy and broad-band photometry. Figure \ref{fig:sample_properties} displays the galaxies on the plane of stellar mass versus \nuvr\ color, as well as the BPT diagram \citep{Baldwin-Phillips-Terlevich-81}. The stellar mass and \nuvr\ color are taken from the NASA Sloan Atlas (NSA)\footnote{http://www.nsatlas.org}, which is a catalog constructed by \cite{Blanton-11} listing physical parameters for more than 640,000 local galaxies based on data from GALEX, SDSS and 2MASS. The relevant emission-line ratios used for the BPT diagram are taken from the MPA/JHU SDSS database\footnote{http://www.mpa-garching.mpg.de/SDSS/DR7/} \citep{Brinchmann-04}. In the BPT diagram, the dashed and dashed-dotted curves are adopted from \citet{Kauffmann-03a} and \cite{Kewley-06} to separate AGN from star forming galaxies. For comparison, we have selected a volume-limited sample from the NSA, consisting of 35,070 galaxies with $r$-band absolute magnitude brighter than -17.2 and redshift between 0.01 and 0.03. Distributions of this sample are shown as grayscale background in both panels. 

Figure \ref{fig:sample_diagram} shows the relations between the three SFH diagnostics: \dindex, \ewhda\ and \ewhae. These parameters are obtained from the SDSS 3\farcs0-fiber spectra, and so they combine to probe the recent SFH of the central 1-2 kpc of the MaNGA galaxies. We take measurements of these parameters from the MPA/JHU database. As before, distributions of the volume-limited NSA sample are plotted as grayscale background. For comparison, we also show predicted relations of the same parameters by the \cite{Bruzual-Charlot-03} models of solar metallicity that follow exponentially declining star formation histories (${\rm SFR}\propto\exp(-t/\tau)$). Solid lines are for continuous star formation decline with $\tau>5\times10^8$yr, and dashed lines for star formation bursts with $\tau<5\times10^8$yr. Different lines represent the different characteristic timescales $\tau$. The H$\alpha$ luminosity is computed 
by converting Lyman continuum photons to H$\alpha$ following \citet[][see equations B2-B4 in their appendix]{Hunter-Elmegreen-04}, adopting the recombination coefficients and H$\alpha$/H$\beta$ ratios from \citet{Hummer-Storey-87}.

Figure~\ref{fig:sample_properties} shows that our sample well represents the general population of galaxies at stellar mass above $\sim10^9$\msolar, covering the full parameter space in both the color-mass diagram and the BPT diagram. Our sample lacks galaxies below $\sim10^9$\msolar, as well as galaxies with lowest [N{\sc ii}]/H$\alpha$, which can both be attributed to the faint limit of $M_i$ adopted for the MaNGA target sample \citep[see][]{Wake-17}. In addition, our sample appears to have a surfeit of the most massive galaxies compared to the volume-limited NSA sample, which is a consequence of the MaNGA sample design which attempts to make the sampling uniform in mass \citep[][]{Wake-17}. From Figure~\ref{fig:sample_diagram} we see that the central regions of the galaxies form relatively tight relations involving all the three diagnostics, which closely follow the regions of the continuous star formation models. 

Following Paper I, we divide the galaxies into two subsamples according to the central \dindex\ measured from the central spaxel in the MaNGA datacube\footnote{Note that the CFS and CQ classification is based on the MaNGA central spaxel, while the parameters used for making Figure~\ref{fig:sample_diagram} come from the SDSS single-fiber spectroscopy. This is why there isn't a sharp break between red and blue symbols in the figure.}: centrally quiescent (CQ; shown in red triangles) galaxies with \dindex$_{\rm cen}>$1.6, and centrally star-forming (CSF; shown in blue circles) galaxies with \dindex$_{\rm cen}\leq$1.6 \citep{Li-15}. As can be seen from Figure \ref{fig:sample_properties},  almost all CSF galaxies are located in the blue cloud with \nuvr$<4$, and the majority of CQ galaxies are located in the red sequence with \nuvr$>5$. However, a large fraction of CQ galaxies fall in the green valley, and some of the high-mass galaxies (\lgmstar$>$10.0) are found even in the blue cloud. This indicates that high mass galaxies may be still forming stars if seen as a whole, even though the galactic center is already quenched. 

%Compared to Paper I, our sample has as large as 160 times of galaxies. And our sample spreads a very large region on both stellar mass versus \nuvr\ plane and BPT diagram, which means our sample includes a large range of galaxy populations: star-forming galaxies, quenched galaxies and galaxies in transitional phase. This large sample of spatial resolved data enables us to study the resolved star formation histories and the underlying quenching mechanisms for different galaxy populations.  

%In the present work, we do not mainly use quantitative measurements of mean stellar age and star formation rate, while we use D$_n$(4000), \ewhda\ and \ewhae\ instead to trace star formation history. Not only the measurements of these three parameters can very well indicate star formation history \citep{Bruzual-Charlot-03, Kauffmann-03b, Kriek-11}, but also they are more model independent in contrast to mean stellar age and star formation rate. In addition, we also use the mean stellar age measurements to verify our results in appendix. 

\subsection{Full spectral fitting}

We perform full spectral fitting for each spaxel in the MaNGA datacubes following the same method as described in Paper I, and obtain two-dimensional maps of \dindex, \ewhda\ and \ewhae\ for each galaxy. We use the empirical spectral fitting code developed by \cite{Li-05}, which utilizes a set of 9 template spectra constructed from observed stellar and galactic spectra using the technique of principle component analysis. The median $\chi^2$ of the spectral fitting is 1.12 for the whole sample, with only 3.7\% spaxels fitted with $\chi^2>3$. We have visually examined these spectra, and find the large $\chi^2$ is caused by multiple reasons but the regions of the \dindex\ and \ewhda\ are indeed well fitted in most cases. The reader is referred to \cite{Li-05} for a detailed description of the construction of the templates and tests on the spectral fitting results. We then measure \dindex\ and \ewhda\ based on the best fitting stellar spectra without correcting the intrinsic dust attenuation, as well as \ewhae\ from the pure emission line component, obtained by subtracting the best-fit stellar component from the observed spectrum. We have fitted the H$\alpha$ emission line with a single Gaussian profile when measuring \ewhae. Examples of spectral fits and maps of the three diagnostics, as well as tests on the methodology, can be found in Paper I. We also use the public spectral fitting code {\tt STARLIGHT} \citep{CidFernandes-05} to derive a stellar mass for each spaxel, which will be used when obtaining mass-weighted parameters.

\section{Results}

\subsection{A single classifier of star formation status}
\label{sec:fq_definition}

Previous studies usually used broad-band colors (e.g. \nuvr) or spectroscopic diagnostics measured at galactic centers (SDSS-based SFR and \dindex) to divide galaxies into star-forming and quiescent subsamples. As shown above, however, the central \dindex\ and global \nuvr\ color provide inconsistent classifications in many cases. Here, we introduce a new parameter to characterize the overall status of star formation in a galaxy based on 2D maps of \dindex\ and \ewhae. The parameter, $f_Q$, is defined as the mass-weighted fraction of spaxels that are quenched within a given galactic radius $R$, \begin{equation}
f_Q(R) = \frac{\displaystyle\sum_{r<R}m_{\ast,i} \times f_i(D_n(4000),EW(H\alpha))}{\displaystyle\sum_{r<R}m_{\ast,i}}.
\end{equation}
Here $m_{\ast,i}$ is the stellar mass of the $i$th spaxel derived with {\tt STARLIGHT}, and $f_i$ determines whether the $i$th spaxel is quenched or not. A spaxel is quenched if D$_n$(4000)$>$1.6 and \ewhae$<$2.0 \AA\ \citep{Geha-12}, so 
$f_i$ can be written as 
\begin{equation}
 f_i=
   \begin{cases}
   1,   &  Dn(4000)_i>1.6\  {\rm and}\  EW(H\alpha)_i<2.0, \\
   0,   &  \rm Otherwise, \\
   \end{cases}
\end{equation}
where \dindex$_i$ and \ewhae$_i$ are the 4000 \AA\ break and H$\alpha$ emission line equivalent width
for the $i$th spaxel. When calculating $f_Q(R)$, we have corrected the inclination effect for each galaxy, based on the minor-to-major axis ratio taken from NSA.

\begin{figure}
    \epsfig{figure=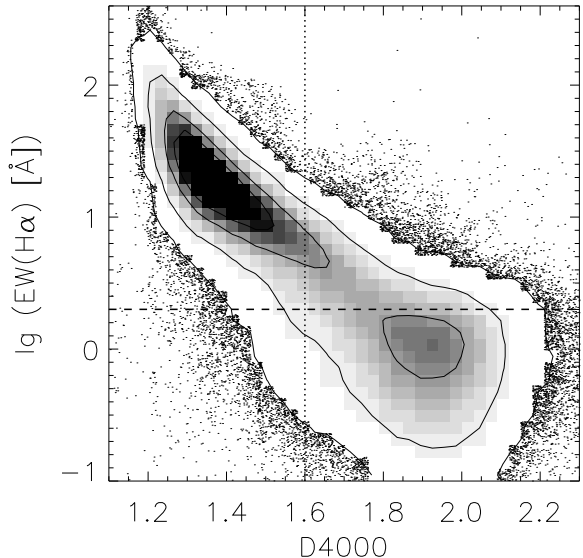,clip=true,width=0.45\textwidth}
  \caption{Distribution of all the individual spaxels in our galaxies on the plane of \dindex\ versus \lgewhae. The horizontal dashed line and the vertical dotted line indicate \ewhae$=$2\AA\ and \dindex$=$1.6, respectively. The grayscale represents the number density of spaxels in linear space on this diagram.}
  \label{fig:halpha_d4000}
\end{figure}

\begin{figure*}
  \begin{center}
    \epsfig{figure=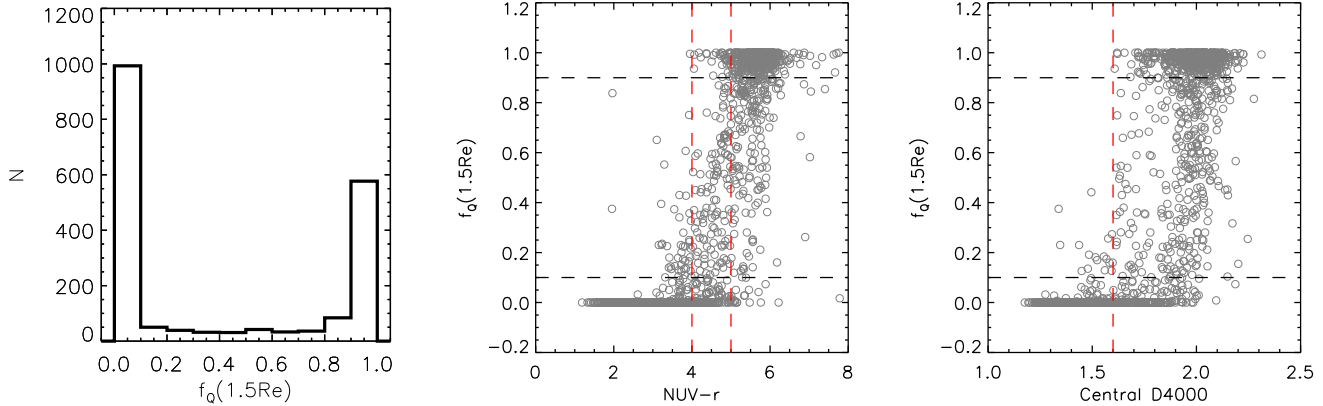,clip=true,width=1.0\textwidth}
  \end{center}
  \caption{Left panel: histogram of $f_Q(1.5R_e)$ for our full sample.
      Middle panel: correlation between \nuvr\ and $f_Q(1.5R_e)$. The two horizontal lines are for $f_Q(1.5R_e)$=0.1 and 0.9, the boundaries used to classify our galaxies into three types. The two vertical lines are \nuvr$=$4 and \nuvr$=$5. Right panel: correlation between central D$_n$4000 and $f_Q(1.5R_e)$. The two horizontal lines are for $f_Q(1.5R_e)$=0.1 and 0.9, while the vertical line indicates D$_n$(4000)$=$1.6.}
  \label{fig:introduce_fq}
\end{figure*}

In this definition, we have used both \dindex\ and \ewhae\ to identify quenched spaxels. Figure \ref{fig:halpha_d4000} displays the distribution of all the spaxels of the sample galaxies on the plane of \dindex\ and \lgewhae. The horizontal dashed line is for \ewhae$=$2 \AA, and the vertical dotted line for \dindex$=$1.6. The spaxels are roughly separated into two regions: star-forming regions with higher \ewhae\ and lower \dindex, and quenched regions with lower \ewhae\ and higher \dindex. This is in agreement with the bimodality of stellar population on kpc scales \citep{Zibetti-17}. As can be seen, the spaxels with \dindex$>1.6$ are mostly, but not always below the line of \ewhae$=2$\AA. A significant fraction of the spaxels with \dindex$>1.6$ can have an \ewhae\ as high as $\sim10$\AA. In previous studies, regions with \ewhae$>$6 \AA\ are usually classified as star forming regions \citep[e.g.][]{Sanchez-14}, while regions  with \ewhae$<$6 \AA\ appear to be diffuse ionized gas, likely ionized by post-AGB stars \citep[e.g.][]{Papaderos-13, Belfiore-17, Zhang-17}. The criterion of \ewhae$<$2 \AA, as an addition to the requirement of \dindex$>1.6$, is probably somewhat conservative, but safe for identifying quenched regions \citep{Geha-12}.

In this paper, we adopt $R=1.5R_e$ when measuring $f_Q$, considering that 1.5$R_e$ is the radius to which both the Primary and Secondary samples are observed with sufficient S/N. In Figure \ref{fig:introduce_fq}, we show the histogram of $f_Q(1.5R_e)$ of our sample (left panel), and the correlation of $f_Q(1.5R_e)$ with both global \nuvr\ (middle panel) and the central \dindex\ (right panel). It is striking that the majority of galaxies can be divided into two extreme cases with either $f_Q<0.1$ or $f_Q>0.9$, with only a small fraction falling in between. Therefore, we classify them into three subsets: star-forming (SF) galaxies with $f_Q(1.5R_e)<0.1$, partly quenched (PQ) galaxies with $0.1\leq f_Q(1.5R_e)<0.9$, and totally quenched (TQ) galaxies with $f_Q(1.5R_e)\geq0.9$. As a result, our sample includes 993 SF galaxies, 347 PQ galaxies and 577 TQ galaxies. Having performed the $1/V_{max}$ correction, we find the SF population dominates the sample ($\sim69.5\%$), TQ galaxies account for $18.4\%$ of the total, and only $\sim12.1\%$ are in the class of PQ.   

Galaxies in the blue cloud with \nuvr$<4$ and those with central \dindex$<1.6$ are mostly classified as SF galaxies with $f_Q(1.5R_e)<0.1$, that is, more than 90\% of the area of these galaxies are forming stars. On the other hand, however, SF galaxies are not always blue or centrally star-forming. Rather, the SF galaxies may have redder colors with \nuvr$>4$ or quenched centers with \dindex$>1.6$. This implies that the central region of these galaxies is quenched, while the outskirts is still forming stars. Galaxies with \nuvr$>4$ or central \dindex$>1.6$ span a full range of $f_Q(1.5R_e)$, suggesting that the global color or central spectroscopy cannot accurately reflect the overall status of the star formation in these galaxies. 

\begin{figure*}
  \begin{center}
    \epsfig{figure=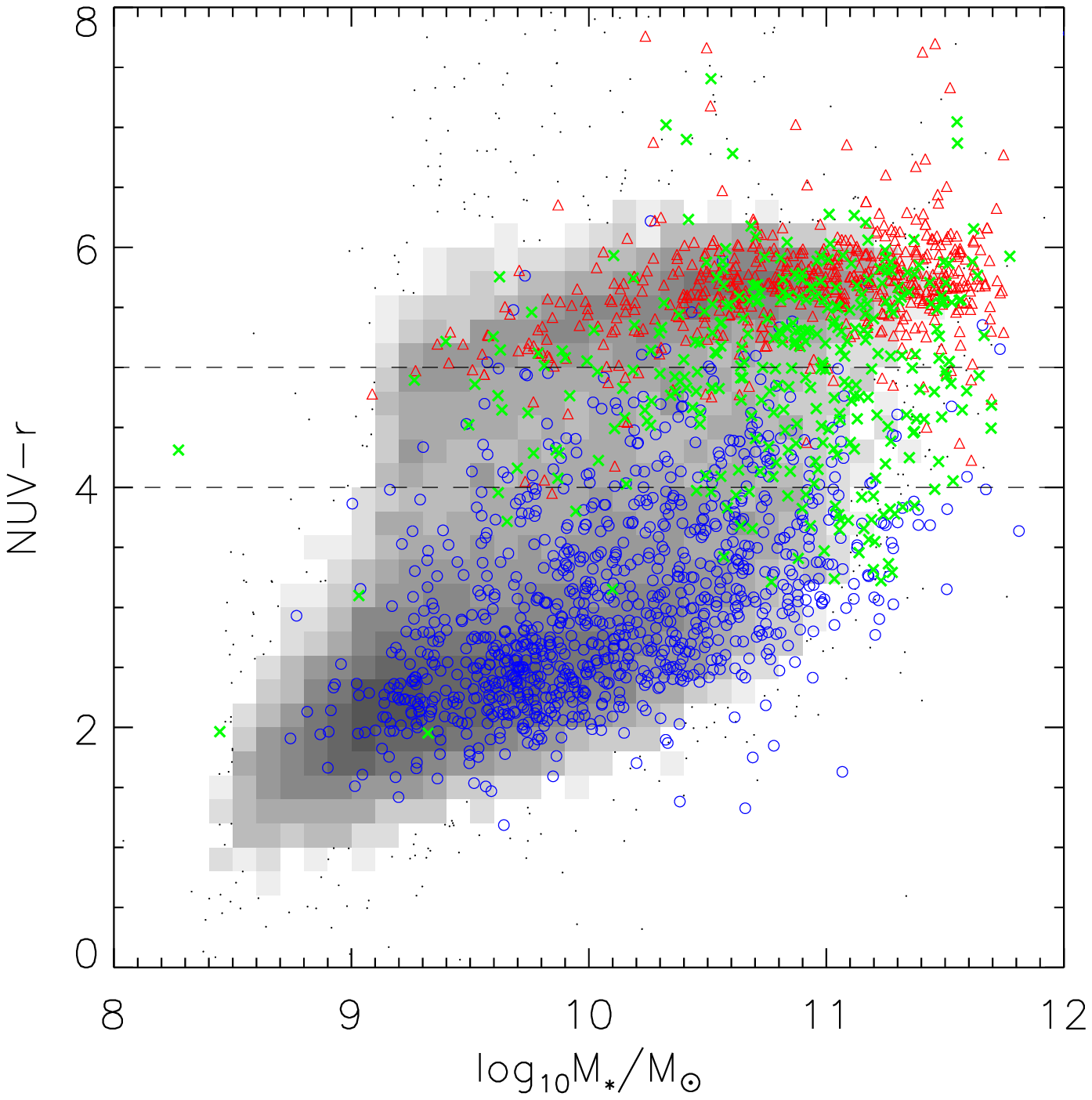,clip=true,width=0.4\textwidth}
    \epsfig{figure=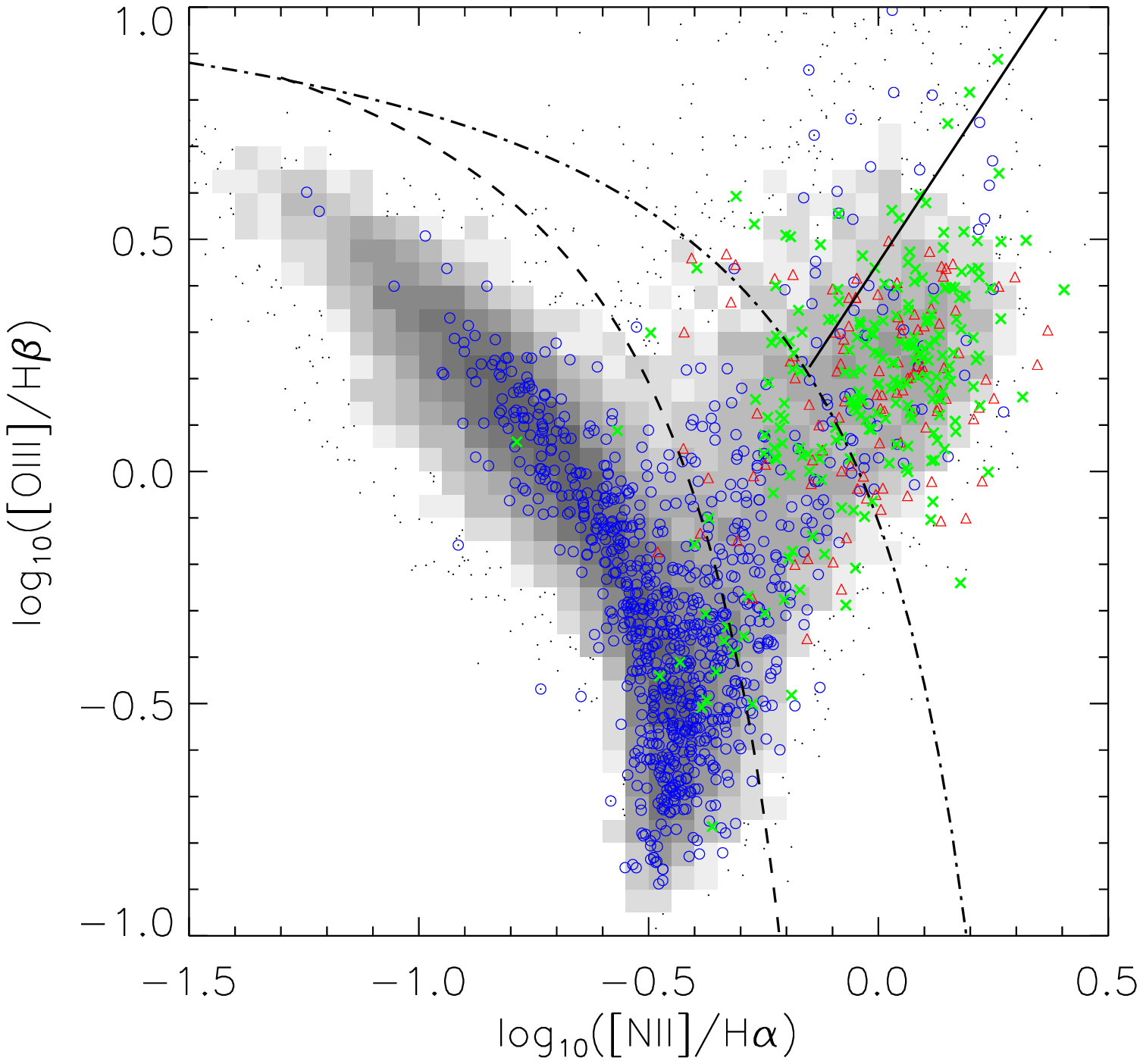,clip=true,width=0.4\textwidth}
  \end{center}
  \caption{ Galaxies are shown on the stellar mass vs. \nuvr\ plane (left panel) and the BPT \citep{Baldwin-Phillips-Terlevich-81} diagram (right panel). The figure is similar to Figure \ref{fig:sample_properties}, but the color of the symbols reflect the overall star formation status of the galaxies, which is fully star-forming (blue), partly quenched (green), or totally quenched (red).
}
  \label{fig:sample_properties_fq}
\end{figure*}

In Figure \ref{fig:sample_properties_fq}, we show the color-mass and BPT diagrams for the sample galaxies again, using the same SDSS-based parameters as in Figure \ref{fig:sample_properties}, but plotting the SF/PQ/TQ galaxies in red/green/blue symbols respectively. SF and TQ galaxies are well separated in the \nuvr\ vs. \mstar\ diagram, while PQ galaxies are broadly distributed over a wide color range above \nuvr$\sim3$. TQ galaxies are mostly redder than \nuvr$=5$, which is the commonly-adopted cut for selecting red-sequence galaxies\footnote{We note that some authors adopted a mass-dependent color division. As can be well expected from Figure~\ref{fig:sample_properties_fq}, our conclusion that TQ galaxies are mostly located in the red sequence would not change if a mass-dependent division was used.}. This demonstrates that the red global color is a necessary, but not sufficient criterion for identifying a fully quenched galaxy. Similarly, blue-cloud galaxies, as usually selected by \nuvr$<4$, are mostly but not purely SF galaxies. The green-valley enclosed by the red and blue divisions, i.e. $4<$\nuvr$<5$, is a mixture of SF and PQ galaxies, with very few TQ galaxies. Interestingly, the relative fractions of the SF and PQ classes in the green valley depends on stellar mass, in the sense that the SF population dominates at low masses (below a few $\times10^{10}$\msolar) and the PQ population dominates at high masses. 

In the BPT diagram, we find the SF galaxies to be distributed over all the different regions, although the majority reside in the expected star-forming region. A small, but significant fraction of the SF galaxies are found in the composition and AGN area. On the other hand, the centrally star-forming galaxies, located below the dividing line of \cite{Kewley-06} in the BPT diagram, are mostly classified as fully star-forming galaxies according to the MaNGA data. This echoes the finding of Paper I, where centrally SF galaxies present weak or no gradients in diagnostics of recent SFH. Furthermore, we find PQ and TQ galaxies are similarly located in the composition and AGN regions, with the majority falling in the region of LINER, or LIER \citep[e.g.][]{Belfiore-15, Belfiore-16, Zhang-17}. This might be suggesting that the two classes of galaxies share similar ionizing sources in the nuclear region.

\subsection{Radial profiles and gradients of the SFH diagnostics}

\begin{figure*}
  \begin{center}
    \epsfig{figure=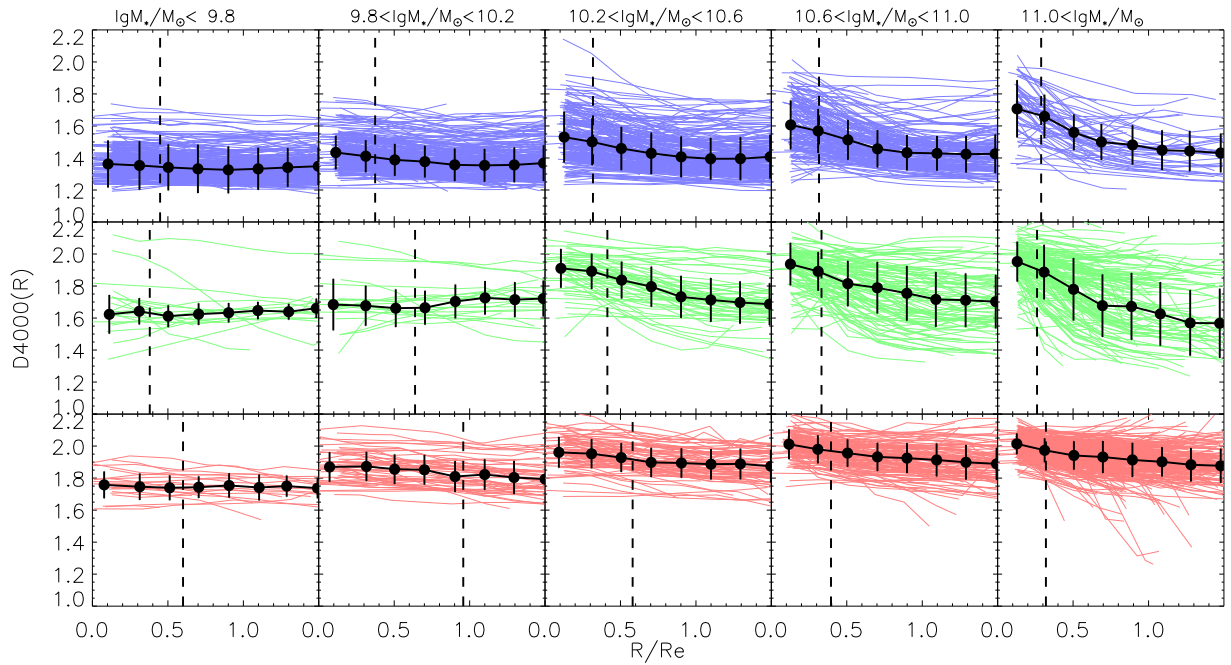,clip=true,width=1.0\textwidth}
  \end{center}
  \caption{Radial profiles of \dindex, plotted as a function of radius scaled by effective radius. Panels from left to right are results for five different stellar mass intervals, as indicated. Panels from top to bottom present results for the three types of galaxies separately: fully star forming (SF) galaxies, partly quenched (PQ) galaxies and totally quenched (TQ) galaxies. Coloured thin lines are for individual galaxies. In each panel, the dots connected with the solid line plot the median profile of all the galaxies in the panel, and error bars indicate the $1\sigma$ scatter of individual galaxies around the median relation. In each panel, the vertical dashed line shows the typical resolution of each subsample.
  }
  \label{fig:D4000_profile}
\end{figure*}

%\begin{figure*}
%  \begin{center}
%    \epsfig{figure=./fig/gline_d4000_fqr15_wei.eps,clip=true,width=1.0\textwidth}
%  \end{center}
%  \caption{Normalized radial profiles of \dindex. Symbols and lines are the same as in the previous figure, except that the radial profile of each galaxy is normalized by subtracting the \dindex\ at 0.5$R_e$.}
%  \label{fig:D4000_profile_mormalized}
%\end{figure*}

\begin{figure*}
  \begin{center}
    \epsfig{figure=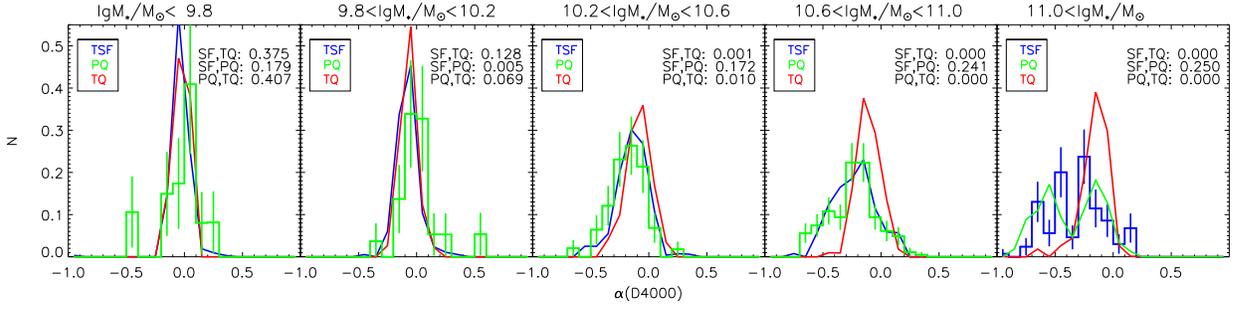,clip=true,width=1.0\textwidth}
  \end{center}
  \caption{Histograms of \dindex\ slope index, $\alpha(D4000)$, for galaxies with different stellar mass (panels from left to right) and at different star formation status, with red/green/blue lines for SF, PQ and TQ galaxies separately. K-S probabilities of the distributions of every two subsamples are indicated in the top-right corner of each panel. In each stellar mass bin, we show the Poisson errors for the subsample with the least number of galaxies. 
  }
  \label{fig:D4000_slope}
\end{figure*}

\begin{figure*}
  \begin{center}
    \epsfig{figure=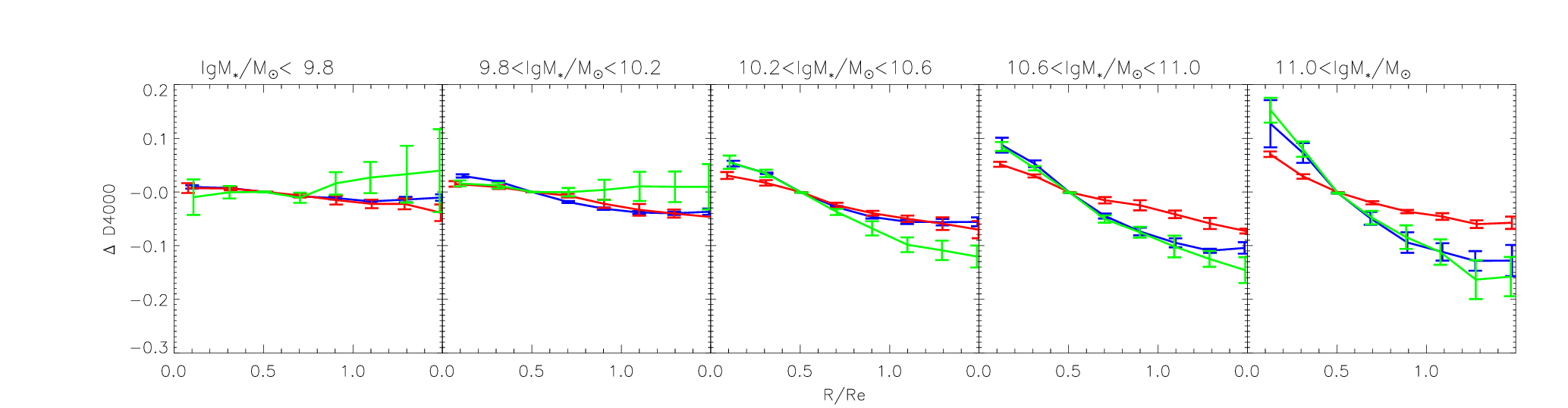,clip=true,width=\textwidth}
    \epsfig{figure=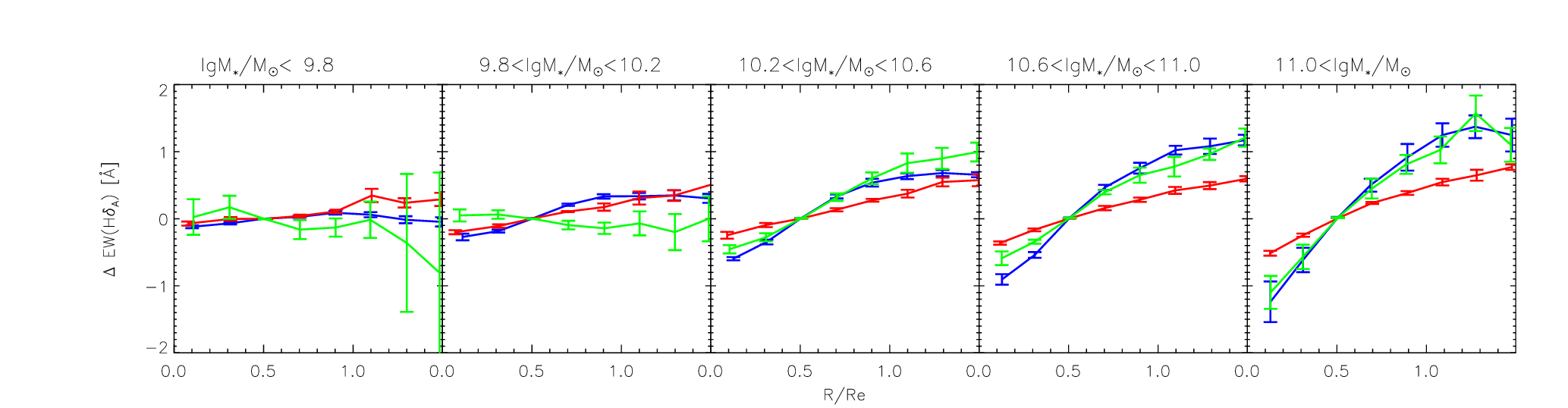,clip=true,width=\textwidth}
    \epsfig{figure=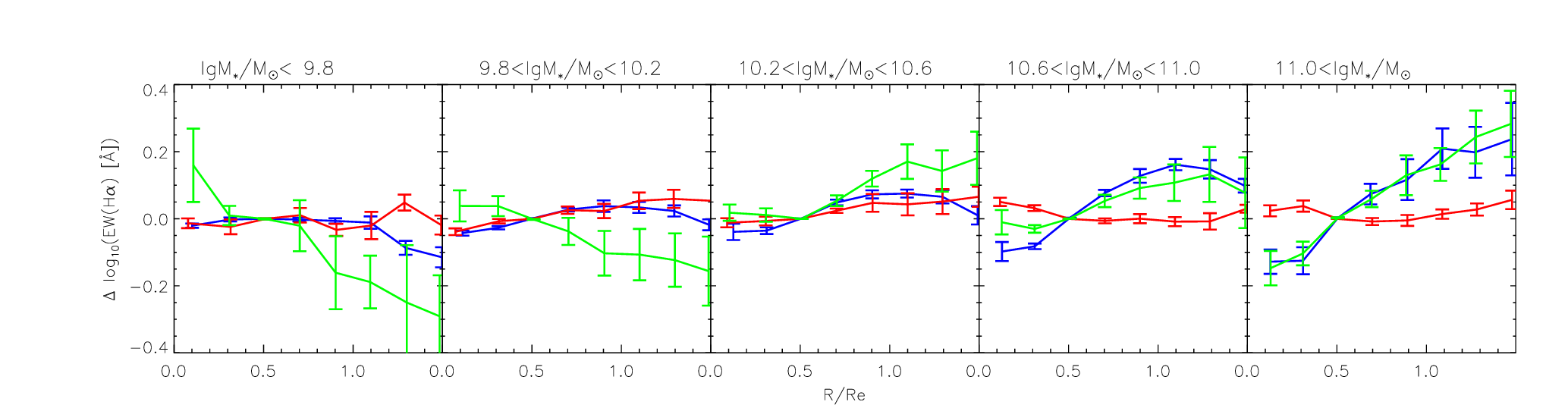,clip=true,width=\textwidth}
  \end{center}
  \caption{ Median radial gradients in \dindex\ (top panels), \ewhda\ (middle panels) and \ewhae\ (bottom panels) for galaxies with different stellar masses (panels from left to right) and at different star formation status (red/green/blue corresponding to SF, PQ and TQ galaxies separately). }
  \label{fig:three_grad}
\end{figure*}

For each galaxy in our sample, we have obtained the radial profiles of the three diagnostic parameters: \dindex, \ewhda, and \ewhae. We take the spaxels with a continuum SNR$>$5 at 5500 \AA\ in the two-dimensional map of a given parameter, and divide them into a set of non-overlapping radial bins with a constant interval of $\Delta(R/R_e)=0.2$. The radial profile of the parameter is then derived from the median value of the spaxels falling in each radial bin. When doing this, we have corrected the inclination effect for each spaxel based on the minor-to-major axis ratio from NSA. 

Figure~\ref{fig:D4000_profile} displays the radial profiles of one of the three diagnostic parameters, \dindex, for all the galaxies in our sample. We divide the galaxies into five stellar mass intervals: \lgmstar$<$9.8, 9.8$\leq$\lgmstar$<$10.2, 10.2$\leq$\lgmstar$<$10.6, 10.6$\leq$\lgmstar$<$11.0 and 11.0$\leq$ \lgmstar. For each stellar mass interval, we present the results for the SF, PQ and TQ galaxies separately. In the figure, panels from left to right correspond to the different mass bins, while panels from top to bottom correspond to the three different galaxy types. In each panel, we show the median radial profile of \dindex\ as black dots connected by the solid line, as well as the 1$\sigma$ scatter around the median profile as error bars. When estimating the median profiles, we have corrected the selection effect of the MaNGA target sample by weighting each galaxy by $1/V_{max}$, as described in the previous section. The profiles are plotted as a function of $R/R_e$, which is the galaxy-concentric radius scaled by the effective radius. 

A number of interesting results can be seen in this figure. First, for a given galaxy type, both the amplitude and the slope of the \dindex\ profile vary with stellar mass, with lower amplitudes and flat slopes at mass below $\sim10^{10}$\msolar\ and higher amplitudes and steeper, negative slopes at higher masses. The trend with mass is more pronounced for SF and PQ galaxies, and the effect is rather weak for TQ galaxies, which present no or very weak gradients in \dindex\ at all masses. Second, at fixed mass, the three types of galaxies present different profiles, but only at intermediate-to-high masses (above $\sim10^{10.2}$\msolar; hereafter, we will refer to galaxies with stellar mass above $\sim10^{10.2}$ as massive galaxies, and those below as less massive galaxies.). At these masses, SF and PQ galaxies show similarly steep profiles with smaller \dindex\ at larger radii, while the TQ galaxies present relatively large \dindex\ over all the radii with no/weak radial variations. At lower masses, different types of galaxies show similarly flat profiles, but different amplitudes, with \dindex\ averaged at $\sim1.4$, $\sim1.6$ and $\sim1.8$ for SF, PQ and TQ galaxies, respectively. 

In most cases, the radial profile of \dindex\ within the effective radius can be simply described by a straight line, consistent with the finding of Paper I. Therefore, following Paper I, we have performed a linear fit to the \dindex\ profile at $R<R_e$ for each galaxy. Figure~\ref{fig:D4000_slope} presents the distributions of the best-fit slope, $\alpha$(\dindex), which is the change in \dindex\ per \Reff, for the SF (blue lines), PQ (green lines) and TQ (red lines) galaxies in the same five mass ranges. In each panel, we perform a Kolmogorov-Smirnov (K-S) test to quantify the significance of the statistical differences between each two distributions, as indicated in the top right corner of each panel. An impressive result from this figure is the similarity between the SF and PQ galaxies, which show very similar distributions of $\alpha$(\dindex) at fixed mass. In addition, both the no/weak gradients of all diagnostics in less massive galaxies and the mass dependence in massive galaxies can be easily identified from the figure. For less massive galaxies, the distribution of $\alpha$(\dindex) is constant at around zero with a narrow width of $\sim0.1$, and the distribution is pretty much the same for the three types of galaxies. For massive galaxies, the SF and PQ galaxies still show similar distributions, but centered at more negative $\alpha$(\dindex) values and with larger widths when compared to the distributions of less massive galaxies. The \dindex\ gradient in TQ galaxies is less dependent on stellar mass, with similarly narrow widths at all masses and slightly stronger (negative) gradients at higher masses. 

Figure~\ref{fig:three_grad} present the normalized median profiles of all the three diagnostic parameters, separately for the five stellar mass intervals. For each galaxy in our sample, we normalize the radial profiles by subtracting the values at 0.5$R_e$, and we obtain the median profile for given stellar mass range and galaxy type in the same way as above. The trends of \dindex\ gradient with both stellar mass and the star formation status as seen in the previous figure are more clearly seen in the top panels of the current figure. Furthermore, as shown in the middle and bottom panels, the other two diagnostic parameters, \ewhda\ and \ewhae, show quite consistent behaviors with the \dindex\ in terms of radial gradients. Less massive galaxies with $M_\ast\lesssim10^{10.2}$\msolar\ show weak gradients, regardless of the overall star formation status. For massive galaxies, the gradients of the diagnostic parameters depend on both stellar mass and the star formation status. TQ galaxies show almost no gradients in \ewhae\ and similarly weak gradients in \dindex\ and \ewhda, while SF and PQ show similar, positive gradients in all the diagnostics with stronger gradients at higher masses.

One may worry that the gradients of our diagnostic parameters might be affected by beam 
smearing effect due to the limited spatial resolution of the MaNGA IFUs. In fact, the great majority of the galaxies in MaNGA Primary and Secondary samples are well resolved \citep{Ibarra-Medel-16}. About 92\% of the sample galaxies have an $R_e$ larger than 2.5$^{\prime\prime}$, the effective spatial resolution of the MaNGA datacubes, and more than 97\% have the 1.5$R_e$ larger than 2.5$^{\prime\prime}$. We have also examined the potential dependence of \dindex\ gradients on galaxy angular size for Primary sample and Secondary sample separately, finding no significant trends. Therefore, beam smearing effect should not have a significant impact on the results.

\subsection{Diagnostic diagrams of Recent SFH}

\begin{figure*}
  \begin{center}
    \epsfig{figure=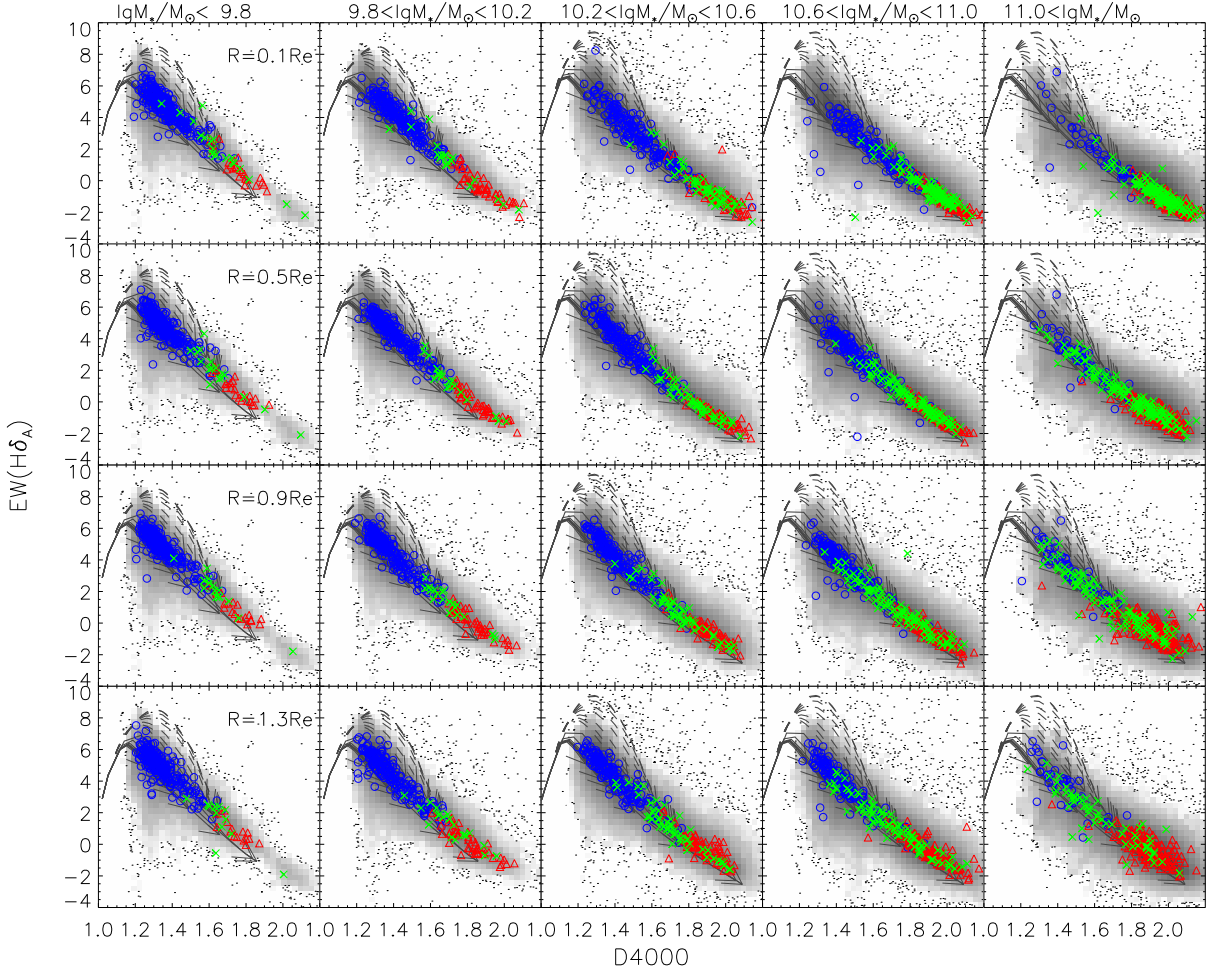,clip=true,width=1.0\textwidth}
  \end{center}
  \caption{Distributions in the plane of \dindex\ and \hda\ indices are shown for different stellar mass intervals (panels from left to right) and different radial bins (panels from top to bottom). The stellar mass ranges are indicated above the top panels, and the center radii of the radial bins are indicated in the left-most panels. Blue/green/red symbols represent the spaxles from the corresponding mass and radial bins. The distribution of all the spaxels from a given stellar mass bin is plotted as grayscale background for comparison. The lines in each panel are predictions of the \cite{Bruzual-Charlot-03} models with either continuously declining star formation (solid lines) or starbursts (dashed lines). We adopt the BC03 models with $Z=0.4Z_{\odot}$ for the lowest two stellar mass bins, and $Z=Z_{\odot}$ for the other stellar mass bins. See the text for details.}
  \label{fig:diagram_one}
\end{figure*}

\begin{figure*}
  \begin{center}
    \epsfig{figure=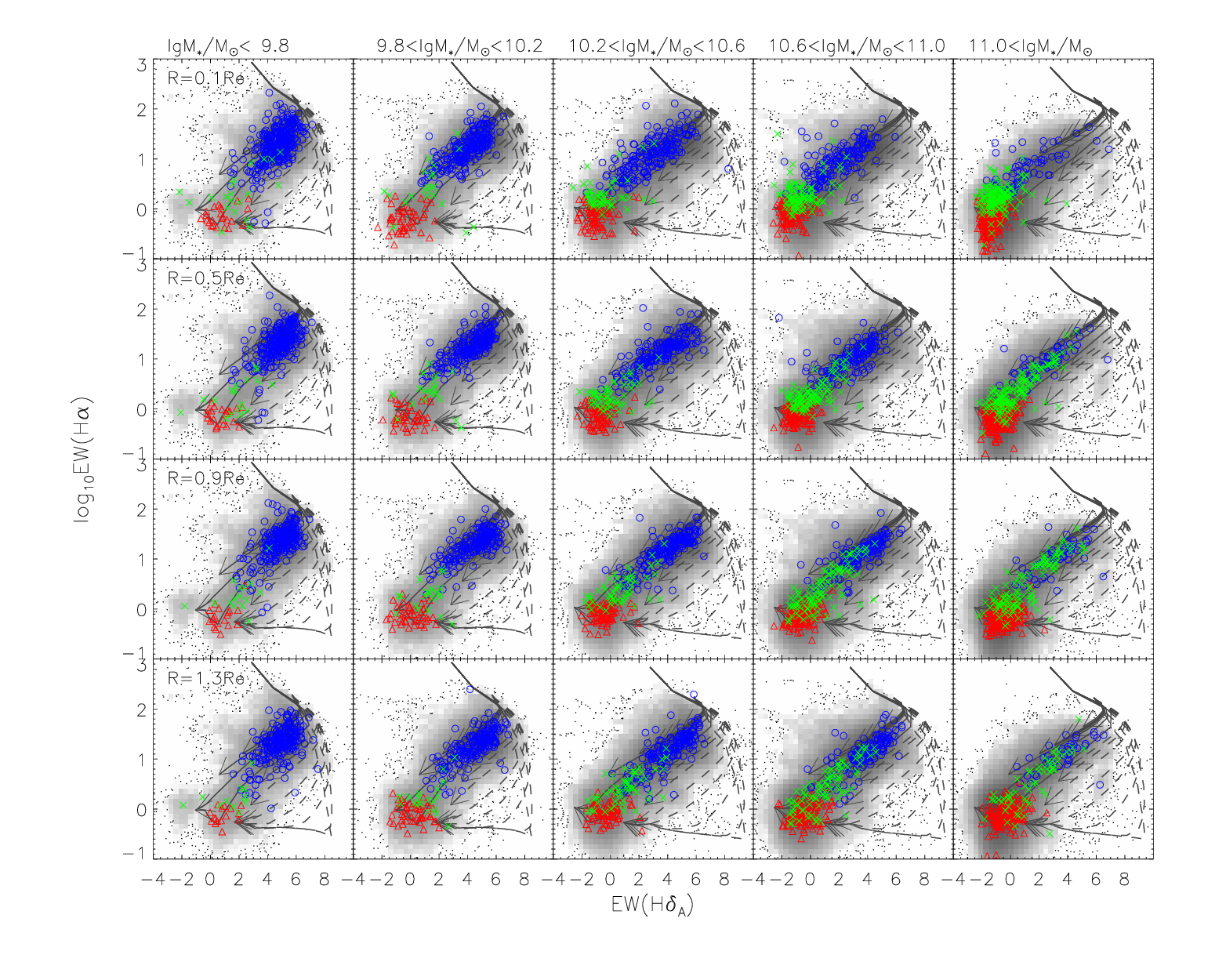,clip=true,width=1.0\textwidth}
  \end{center}
  \caption{Same as Figure \ref{fig:diagram_one}, but showing distributions in the plane of \ewhda\ and \lgewhae.}
  \label{fig:diagram_two}
\end{figure*}

\begin{figure*}
 \begin{center}
   \epsfig{figure=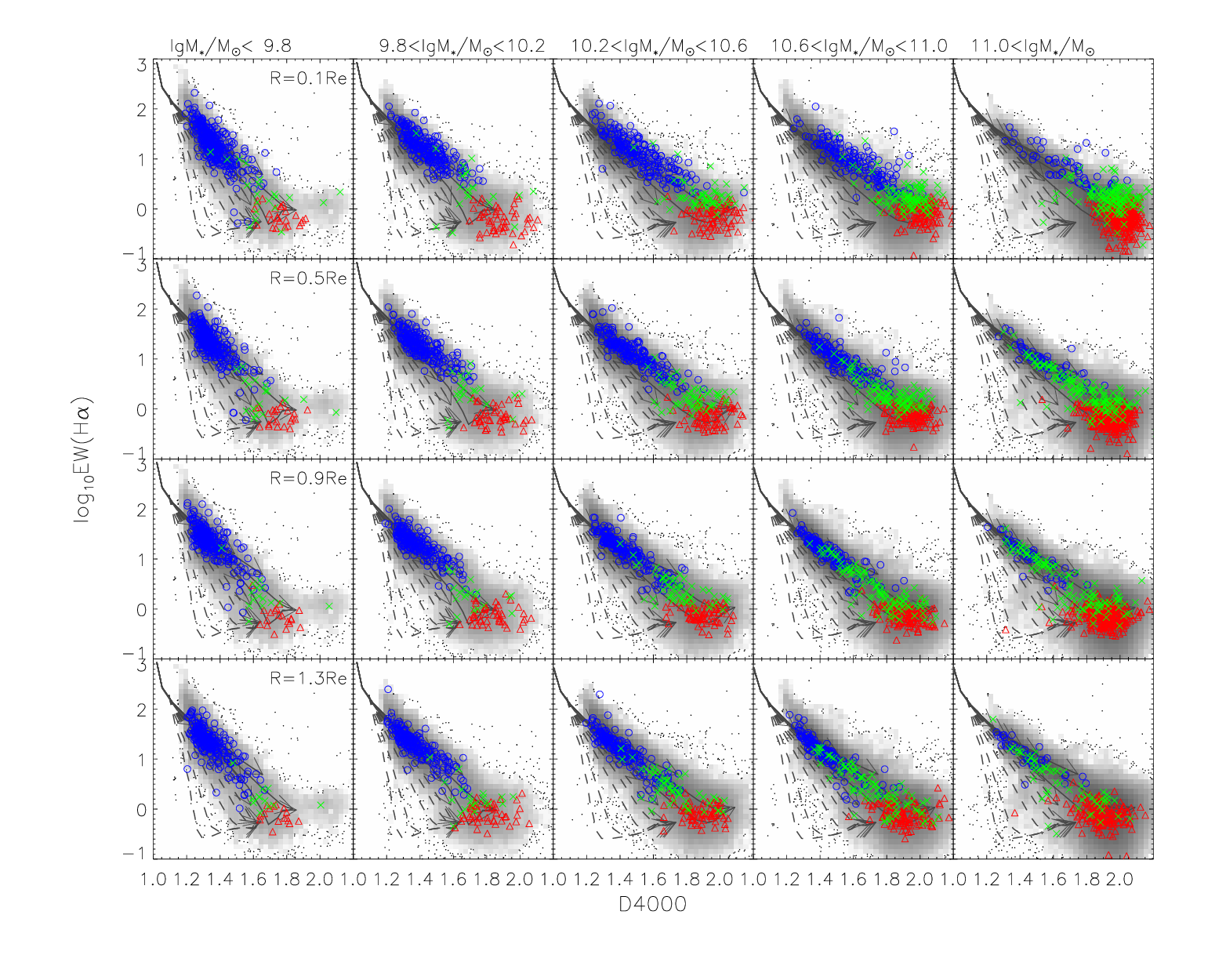,clip=true,width=1.0\textwidth}
  \end{center}
  \caption{Same as Figure \ref{fig:diagram_one}, but showing distributions in the plane of \dindex\ and \lgewhae.} 
 \label{fig:diagram_three}
\end{figure*}

In the previous subsection, we have examined the radial profiles of the three diagnostics. In this subsection we jointly analyze the different diagnostics, by showing their radial profiles on three diagrams of \dindex\ vs. \ewhda, \ewhda\ vs. \ewhae, and \ewhae\ vs. \dindex, respectively. To this end, we take the data points of each radial profile at four different radii: 0.1\Reff, 0.5\Reff, 0.9\Reff\ and 1.3\Reff. For each radius, we then show the distribution of the galaxies on the three diagnostic diagrams. The results are shown in  Figures~\ref{fig:diagram_one}, \ref{fig:diagram_two} and \ref{fig:diagram_three}. Panels from left to right correspond to the five stellar mass ranges, while panels from top to bottom correspond to the four radii. Thus each panel displays a diagram for a given stellar mass bin in a given radial bin. In each panel, the blue circles, green crosses and red triangles represent the SF, PQ and TQ galaxies falling in the stellar mass range. For comparison, we plot all the spaxels of galaxies in the corresponding stellar mass range which have a continuum SNR$>$5 at 5500 \AA, as the background grayscale distribution in each panel. The BC03 models with continuously declining star formation and bursting star formation are plotted as the solid and dashed lines. We note that we have adopted BC03 models of different metallicities for different mass bins, considering the known metallicity dependence on mass \citep{Lequeux-79, Garnett-Shields-87, Tremonti-04,Gallazzi-05,Panter-08, Sanchez-13}.
We use models of $Z=0.4Z_{\odot}$ for less massive galaxies in the two lowest stellar mass bins, and $Z=Z_{\odot}$ for massive galaxies in the other three mass bins.

A general result of these figures is that the distributions of galaxies on these diagrams broadly follow the continuous star formation models, although a small fraction of individual spaxels may extend to regions of starbursts. This is true for all the masses and radii, and also independent of the overall star formation status of the galaxies. A galaxy may be located differently in the diagrams, depending on mass, radius and classification, but the location is confined to the regions of continuous star formation models. For instance, in Figure~\ref{fig:diagram_one}, we see that the PQ galaxies (green crosses) in the top right-most panel (\lgmstar$>11$, $R=0.1$\Reff) are mostly located in the quenched region with largest \dindex\ and smallest \ewhda, and they are moving --- along the tight sequence of continuous star formation models --- to more star-forming regions with smaller \dindex\ and larger \ewhda\ when one goes to larger radii. This result reflects the radial profile of PQ galaxies at fixed mass, as already presented in previous figures. However, the tight sequence on the \dindex-\ewhda\ diagram can only be clearly seen when the diagnostics are jointly examined. It is striking to see the same tight sequence on the diagnostic diagram to hold regardless of mass, radius and galaxy type. But one should keep in mind that we have excluded mergers, and irregular and disturbed galaxies from our sample. The result suggests that, as pointed out in Paper I, the star formation cessation in a galaxy must be a smooth long-term process, likely governed by a common set of drivers, and starburst activities happen rarely in galaxies with regular morphologies. We emphasize that the lack of starburst spaxels are not due to problematic spectral fitting of post-starburst galaxies. We have examined some known post-starburst galaxies in our sample, finding their spectra to be reasonably well fitted.

The distributions of the galaxies on the other two diagrams, both involving \ewhae, are not as tight as those in the \dindex\ vs. \ewhda\ diagram. However, even with a more scattered distribution, the galaxies are still barely found in the regions of starburst models, reinforcing the conclusion that the star formation history in individual spaxels is consistent with the continuous star formation model. Furthermore, we find that the PQ and TQ classes are better separated on the \ewhae-related diagrams, with TQ galaxies located mostly below \lgewhae$\sim0$ and PQ galaxies above it. 

We note that there are outliers (though small in number) in some panels, especially the panels for $R=0.1$\Reff\ and high mass bins in Figure~\ref{fig:diagram_one} and~\ref{fig:diagram_two}. It is interesting that the outliers in these panels are no longer discrepant when one moves to larger radii, implying that these outliers are mostly caused by the inner regions of massive galaxies. We have visually inspected the spectra of the outliers, finding their emission lines to present a very broad component, thus indicating the presence of a type-I AGN. Special treatments and analyses of AGN are beyond the scope of this paper, and the interested reader is referred to \cite{Belfiore-16} and \cite{Zhang-17} for detailed studies of AGN and diffuse ionized gas emission in MaNGA galaxies. 

\subsection{The central \dindex\ versus $\alpha$(\dindex) relation}
\label{sec:d4000_in_out}

\begin{figure*}
  \begin{center}
    \epsfig{figure=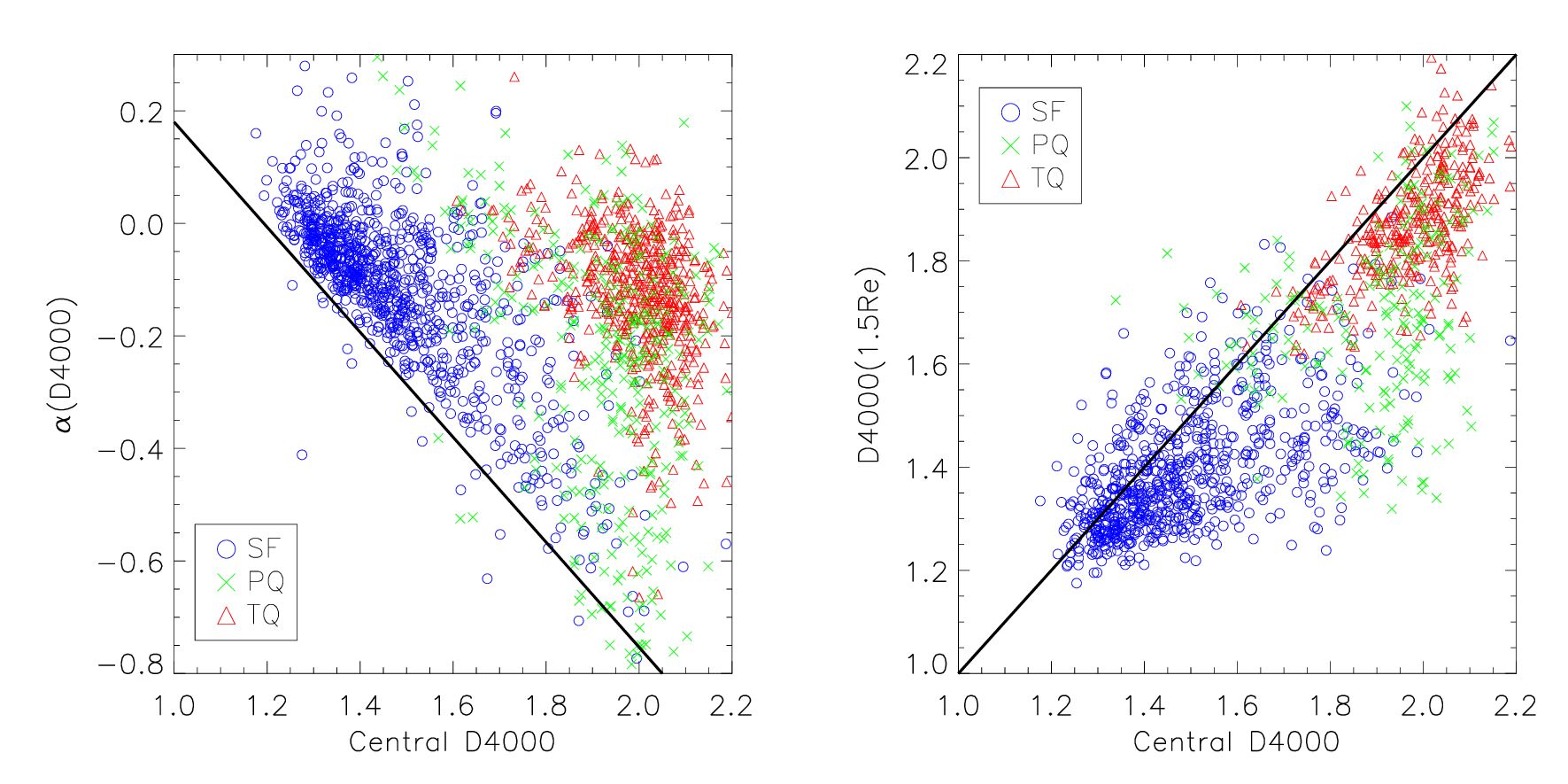,clip=true,width=1.0\textwidth}
  \end{center}
  \caption{Galaxies are plotted on the plane of central \dindex\ versus the slope index of the \dindex\ profile (left-hand panel), and the plane of central \dindex\ versus the \dindex\ at 1.5\Reff\ (right-hand panel). The solid line in the left panel is an arbitrary line to emphasize the nearly-linear sequence of SF galaxies in this diagram, and the line in the right panel represents the 1:1 relation. Blue circles, green crosses and red triangles are for the subsets of fully star-forming, partly quenched and totally quenched galaxies, separately. See the text for detailed description of the classification method.}
  \label{fig:D4000_in_out}
\end{figure*}

\begin{figure*}
  \begin{center}
    \epsfig{figure=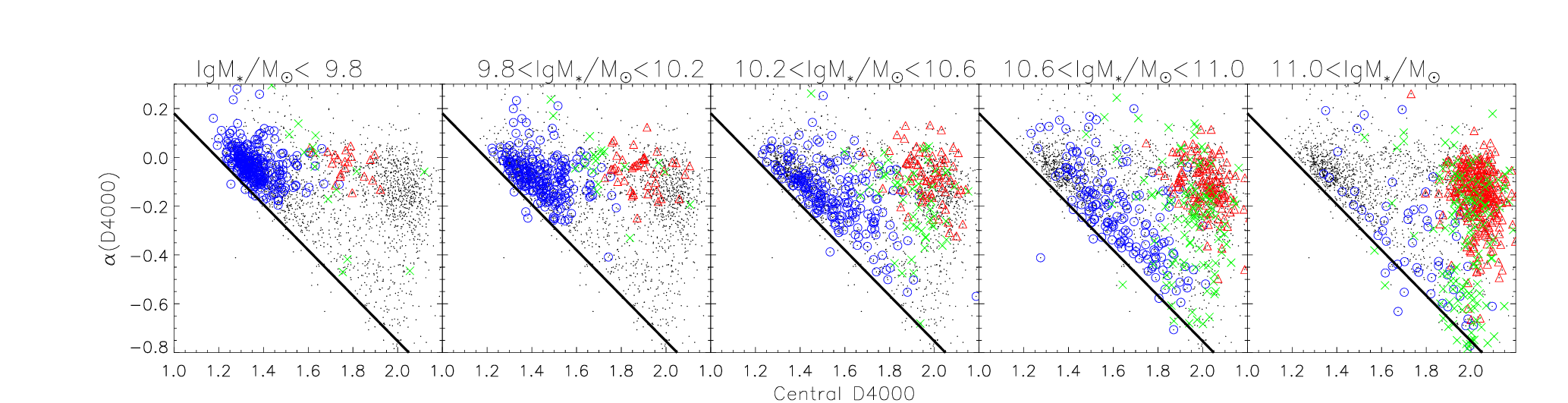,clip=true,width=1.0\textwidth}
    \epsfig{figure=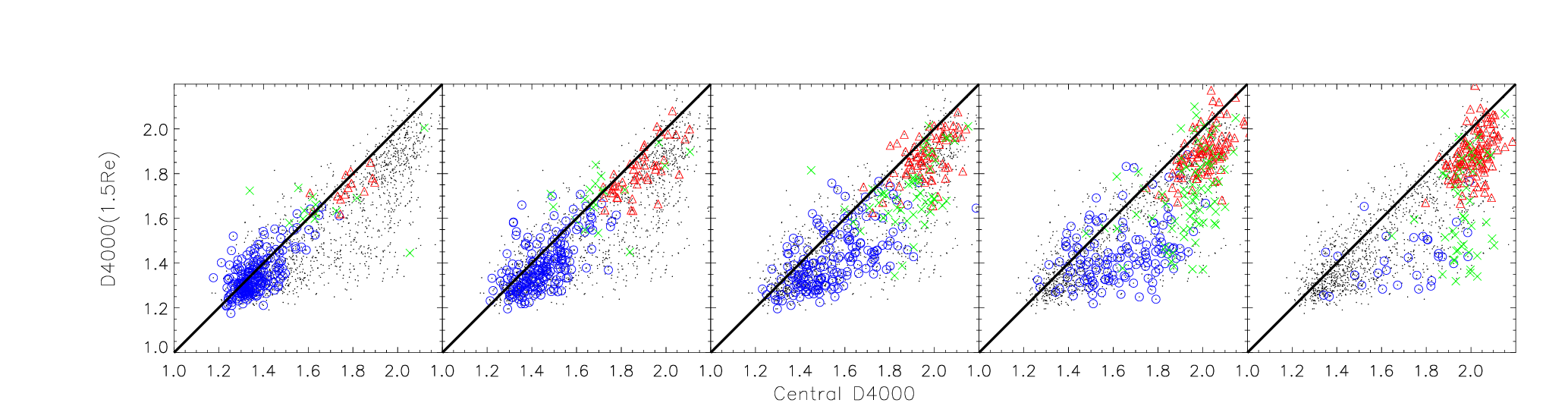,clip=true,width=1.0\textwidth}
  \end{center}
  \caption{The top (bottom) panels are the same as the left-hand (right-hand) panel in the previous figure, with the different panels for five stellar mass intervals.}
  \label{fig:D4000_in_out_mass}
\end{figure*}

In the previous section we have examined the recent SFH for individual spaxels at different radii within the galaxies, finding them to uniformly follow the BC03 models with continuously declining star formation rate, a result which is interestingly independent of stellar mass, radius and the overall star formation status of galaxies. In this subsection we make direct comparisons between the central and outer regions of the galaxies, in order to understand whether there is any evolution relation across the galaxy. We concentrate on one of the three diagnostics, \dindex, for simplicity. Figure \ref{fig:D4000_in_out} compares the \dindex\ measured in the central spaxel (\dindex$_{\rm cen}$) with both the \dindex\ gradient ($\alpha$(\dindex), left panel), and the \dindex\ measured at 1.5\Reff\ (\dindex$_{\rm 1.5Re}$, right panel). In both panels, galaxies are divided into SF (blue circles), PQ (green crosses) and TQ (red triangles) populations.

As can be seen, SF and TQ populations can be well separated in both panels, while PQ galaxies are largely found in between, with substantial overlap with the TQ population. Consistent with the radial profiles shown in previous figures, the majority of the SF galaxies have the lowest \dindex\ in the center, as well as weak/no gradients. Therefore, they are located in the left-upper part on the \dindex$_{\rm cen}$ vs. $\alpha$(\dindex) plane and the left-lower part on the \dindex$_{\rm cen}$ vs. \dindex$_{\rm 1.5Re}$ plane, where \dindex$_{\rm cen}<1.6$,  $\alpha$(\dindex)$>-0.2$ and \dindex$_{\rm 1.5Re}<1.6$. In addition, it is interesting to see that a significant fraction of SF galaxies present strong gradients and high \dindex\ in their centers (\dindex$>1.6$), but low \dindex\ at $1.5$\Reff. They are classified as SF because most of their spaxels fulfill the criteria for SF regions, and only the central region is quenched. As a whole, the SF galaxies form a nearly linear sequence on the \dindex$_{\rm cen}$ vs. $\alpha$(\dindex) plane. The TQ galaxies are mostly located in the top right corner in both panels. They span a relatively narrow range in \dindex\ at both the galactic center and 1.5\Reff, with \dindex$>1.8$ and $\alpha$(\dindex)$>-0.5$ in most cases. The distribution of PQ galaxies is more scattering when compared to the other two classes, covering both the intermediate region between the SF and TQ populations and the whole region of the TQ population, but with little overlap with the SF population. 

These results are in good agreement with the ``inside-out'' picture of star formation cessation, in which the star formation gets shutdown in the galactic center before propagating to larger and larger radii. In this picture, a fully SF galaxy starting with a flat \dindex\ distribution with \dindex\ constant at the lowest values (\dindex$\sim1.2$), will have larger \dindex$_{\rm cen}$ and steeper slopes as the star formation cessation happens from the center outwards. Initially located at the upper-left corner in the the \dindex$_{\rm cen}$ vs. $\alpha$(\dindex) plane, the galaxy appear to move along the SF sequence towards the lower-right corner, where it deviates from the SF sequence and moves upward with central \dindex\ saturated at \dindex$_{\rm cen}\sim2$ and $\alpha$(\dindex) increasing from $\sim-0.8$ (steepest slopes) up to around zero (flat slopes). The same picture can also be seen in the right panel, where the galaxy starts from the lower-left corner with the same \dindex\ at $R=0$ and $1.5$\Reff, evolves to the PQ phase which shows the largest deviation from the $1:1$ relation because of the strongest gradients in \dindex, and comes back to the $1:1$ relation when the galaxy gets totally quenched, thus showing highest \dindex\ in both the center and the outskirts. The PQ galaxies are in the transition phase between the SF and TQ populations, given their locations on both diagrams.

\begin{figure*}
  \begin{center}
  \epsfig{figure=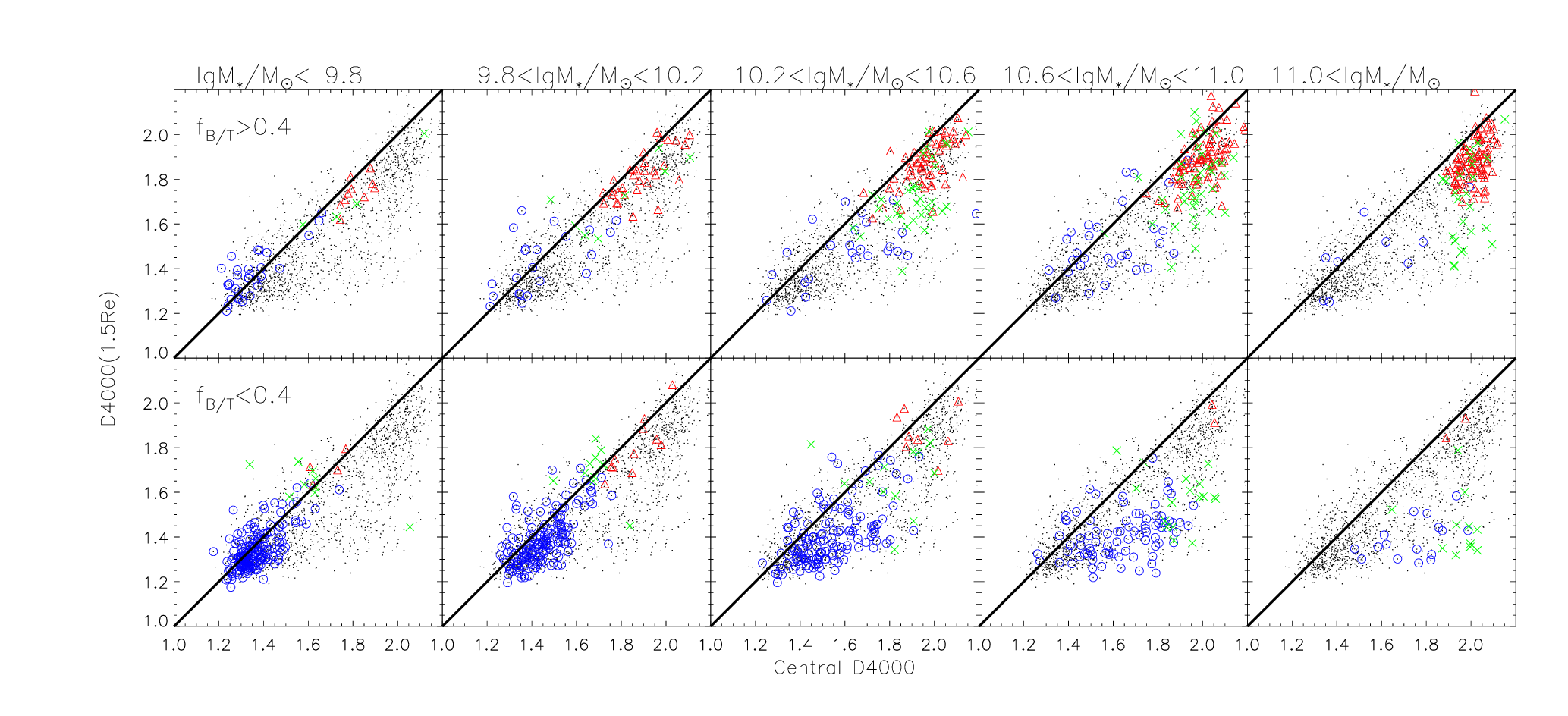,clip=true, width=1.0\textwidth}
  \end{center}
  \caption{Correlation between central 4000 \AA\ break and the 4000 \AA\ break at 1.5\Reff. Panels from left to right correspond to different stellar mass ranges as indicated above the top panels. The top panels shows results for galaxies with $f_{B/T}>0.4$, and the bottom panels for those with $f_{B/T}<0.4$.
  Colors and symbols are the same as in Figure \ref{fig:D4000_in_out}.}
  \label{fig:dependence_bulge}
\end{figure*}

\begin{figure*}
  \begin{center}
  \epsfig{figure=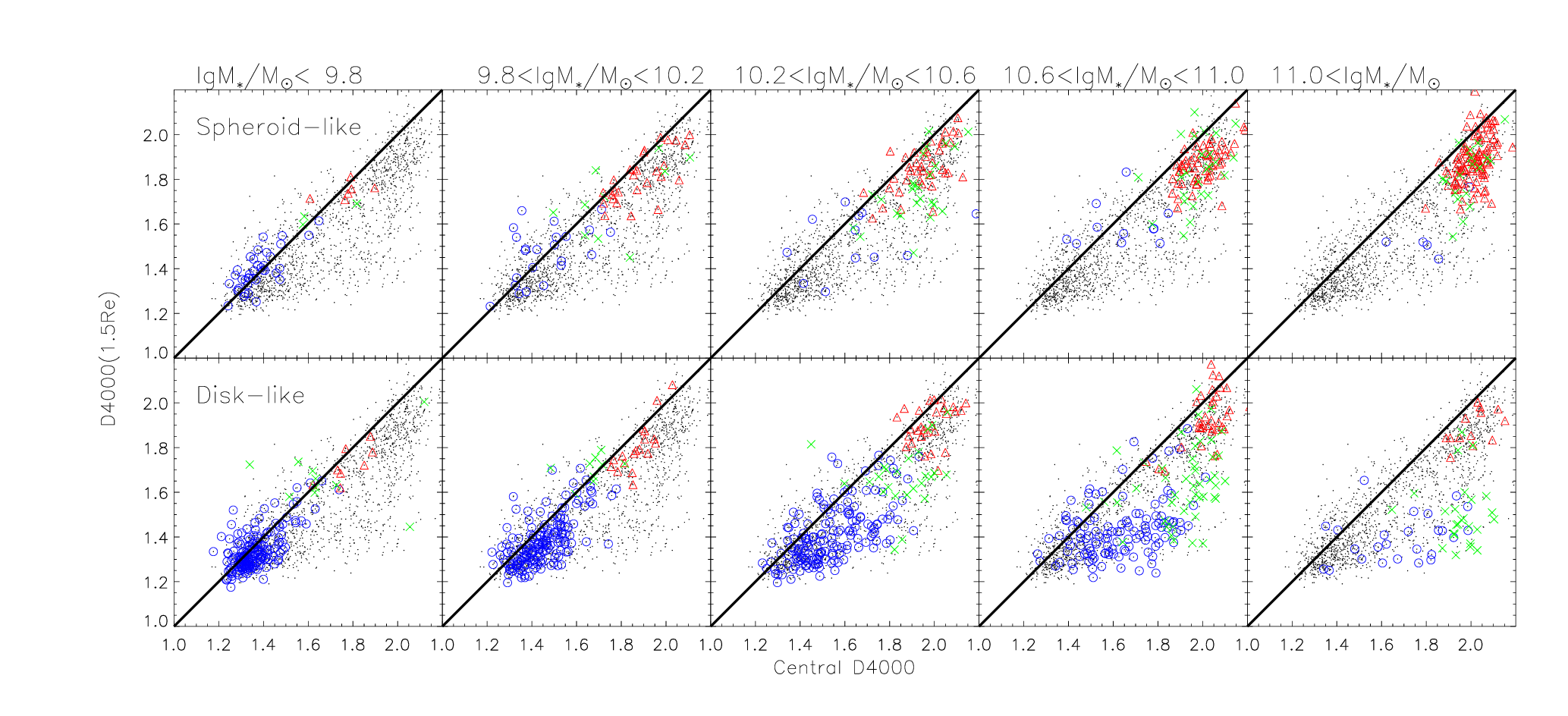,clip=true, width=1.0\textwidth}
  \end{center}
  \caption{Correlation between central 4000 \AA\ break and the 4000 \AA\ break at 1.5$R_e$. Panels from left to right correspond to different stellar mass ranges as indicated above the top panels. The top panels shows results for galaxies with spheroid-like morphology, and the bottom panels for those with disk-like morphology. Colors and symbols are the same as Figure \ref{fig:D4000_in_out}.}
  \label{fig:dependence_mor}
\end{figure*}

Although the two panels generally illustrate the ``inside-out'' star formation cessation process, it is apparent that some individual galaxies do not simply follow this sequence. For instance, some SF galaxies appear to deviate from the SF sequence with a rather flat profile, falling in between the SF sequence and the region of TQ galaxies in the left panel. In the right panel, these galaxies follow the $1:1$ relation even when their central \dindex\ exceeds 1.6. These galaxies are obviously not following the ``inside-out'' picture. In order to understand what drives the diversity in these panels, we have examined the possible dependence on stellar mass, morphology and structural parameters. The results are shown in Figures~\ref{fig:D4000_in_out_mass}, \ref{fig:dependence_bulge} and~\ref{fig:dependence_mor}. 

In Figure~\ref{fig:D4000_in_out_mass}, we repeat the two panels of the previous figure, but for the five stellar mass bins separately. An immediate result from this figure is that the ``inside-out'' process as described above is seen only for massive galaxies in the three high-mass samples with mass above $10^{10.2}$\msolar, with the trend becoming stronger at higher masses. In these mass bins, the galaxies are clearly offset from the one-to-one line, with the deviation increasing gradually with increasing stellar mass. For less massive galaxies, we see no/weak signature of ``inside-out'' in the sense that the galaxies show fairly flat \dindex\ profiles although the central \dindex\ span a full range of $1.2<$\dindex$<2$. We note that there are also a small number of massive SF galaxies with \dindex$_{\rm cen}<$\dindex$_{\rm 1.5Re}$, suggesting that they have different star formation cessation picture from normal massive SF galaxies.  These galaxies have slightly positive gradients in \dindex, and negative gradients in \lgewhae. \cite{Wang-17} used the public MaNGA data from SDSS/DR13 and examined these galaxies, finding them to have smaller size, higher concentration, higher SFR and higher gas-phase metallicity when compared to normal SF galaxies of similar masses. In addition, their median surface mass density profile falls in between the profiles of normal SF galaxies and quiescent galaxies with the same stellar mass distributions, indicating that these galaxies are likely in the transition phase from normal SF galaxies to quiescent galaxies, with rapid on-going central stellar mass assembly, probably related to bulge growth. 

In Figures~\ref{fig:dependence_bulge} and~\ref{fig:dependence_mor}, we further examine the dependence of the diagnostic diagrams on both the bulge-to-total ratio ($f_{\rm B/T}$) and morphological type (spheroid-like vs. disk-like), but showing the results only for the plane of \dindex$_{\rm cen}$ and \dindex$_{\rm 1.5Re}$ for simplicity. The bulge mass fractions are taken from \citet{Simard-11}, obtained by performing photometric decomposition modelling on the SDSS $r$-band image. The morphological classification is done by visually examining the $r$-band image of the sample galaxies. We see quite similar results from both figures in the following three aspects. First, the radial variations in \dindex, as seen at \lgmstar$>10.2$ in Figure~\ref{fig:D4000_in_out_mass}, are dominated by disk-like galaxies in the class of SF or PQ, which have small B/T luminosity ratios. We note that this trend might be driven by the lack of SF spheroid-like galaxies. However, given the small number of SF spheroid-like galaxies at fixed \mstar, we are unable to reliably tell whether the distribution of SF disk-like on the diagram differs significantly from that of the SF spheroid-like galaxies. We will come back to this point in future when larger samples are available. Second, the TQ galaxies are mostly spheroid-like and have large B/T ratios, consistent with previous studies that a dense bulge is usually found in a quenched galaxy \citep[e.g.][]{Fang-13, Bluck-14}. Third, these effects are stronger at higher stellar masses. Stellar mass is apparently a primary driver for the star formation cessation process in local galaxies. These results imply that the presence of a central massive object, such as a bulge, is unlikely to play a driving role in quenching. In fact, an additional analysis shows that the significant deviation of the massive galaxies from the 1:1 relation as seen in Figure~\ref{fig:dependence_bulge} is still hold even when we restrict ourselves to the galaxies with $f_{B/T}<0.2$. On the other hand, the absence of a bulge seems to be even required in order to have a large radial gradient, which is an observational signature of the inside-out cessation process. However, this result should not be overemphasized due to the small number of star-forming galaxies with $f_{B/T}>0.4$. A spheroid-like morphology or a large $f_{B/T}$ is associated mostly with galaxies at the TQ stage, implying that the growth of a central massive object is likely a consequence of the star formation cessation, and probably plays crucial roles only at the late stage of the quenching process.

\section{Discussion}

\subsection{Comparison with previous results}

Paper I investigated maps and radial profiles of \dindex, \ewhda\ and \ewhae\ for a preliminary sample of 12 galaxies observed in the MaNGA prototype run (P-MaNGA), finding the individual spaxels of the P-MaNGA galaxies to broadly follow models of continuously declining star formation and nicely form a tight sequence on the \dindex-\ewhda\ diagram. In this work, we have extended the analysis of Paper I by using a much larger sample of 1917 galaxies from the MaNGA MPL5, and find that those suggestive results are confirmed at much greater significance (see Figure \ref{fig:diagram_one}, Figure \ref{fig:diagram_two} and Figure \ref{fig:diagram_three}). This indicates that the growth and death of galaxies must be a smooth process, and that starbursts happen rarely in galaxies with regular morphology. 

Paper I classified galaxies as either centrally star-forming or centrally quiescent according to the star formation status of the central region, and found the two classes present distinct radial profiles in the diagnostic parameters. In this work, we found that the central region alone can not accurately determine the overall star formation status of a galaxy. A galaxy with a quenched center could be forming stars in the outskirts, thus a partly quenched galaxy, or it could be a totally quenched galaxy with no star formation across the whole area. We have proposed a single classifier, $f_Q(R<1.5R_e)$, based on the 2D map of both \dindex\ and \ewhae, and made comparisons with the conventional classifiers such as global \nuvr\ color and central \dindex. 

%The less massive and massive galaxies show very different behaviors on the radial profiles of three diagnostics, in the sense that less massive galaxies show no or very weak radial variations of these diagnostics, while massive galaxies show very significant gradients, especially for SF and PQ populations. These radial gradients are increasing with increasing stellar mass. This suggests that galaxy evolution from star-forming to quenched population is regulated by different physical mechanism for galaxies with different stellar mass. For massive galaxies, our result is very well consistent with the ``inside-out'' picture of star formation cessation, where the star formation cessation first occurs in galactic center and slowly propagates out to larger radii. However, this is not the case for less massive galaxies. The radial profiles of diagnostics for less massive galaxies show that the star formation cessation occurs simultaneously for both central regions and outer regions. 

Resolved stellar populations have been studied in recent years, based on multi-wavelength broad-band photometry, long-slit spectroscopy, and IFU spectroscopy for nearby galaxies \citep{Perez-13, GonzalezDelgado-14, GonzalezDelgado-15, Sanchez-14}. Using 105 galaxies from CALIFA, \citet{Perez-13} studied the spatially resolved history of their stellar mass assembly by applying the fossil record method, revealing an inside-out growth picture for massive galaxies and a transition to outside-in growth for low mass galaxies (\lgmstar$<$10.0). Following this work, \cite{Ibarra-Medel-16} reconstructed the radial stellar mass growth histories of a large sample of galaxies from MaNGA. They also found an ``inside-out'' stellar mass growth picture for massive galaxies, and this picture is more pronounced in blue/star-forming/late-type galaxies than in red/quiescent/early-type galaxies. In contrast, this picture does not hold for less massive galaxies with stellar mass below $\sim$10$^{10}$M$_{\odot}$. In this work, we have concentrated our investigation on the star formation cessation process rather than the growth process, by using the three diagnostics of recent SFH. We find the ``inside-out''  star formation cessation picture is true only for massive galaxies. For less massive galaxies, we find the recent SFH to be fairly uniform across the galaxy, with no/weak radial gradients. Therefore, these new results support neither the simple ``inside-out'', nor the ``outside-in'' picture for less massive galaxies.

\cite{Zheng-17} estimated the gradients in both stellar age and metallicity by applying the {\tt STARLIGHT} software to an earlier sample from MaNGA, finding significantly negative gradients in age and weak gradients in metallicity. Using a similar MaNGA sample, \cite{Goddard-17} found negative gradients in stellar metallicity for late-type galaxies, with a stronger effect at higher masses. In this work we have taken a less model-dependent approach by focusing our analysis on the three diagnostic parameters. For comparison, we present an analysis of stellar age and metallicity profiles in the Appendix, obtained by performing full spectral fitting to MaNGA datacubes using {\tt STARLIGHT}. Our results are consistent with what \cite{Zheng-17} found. The stellar age gradients obtained from {\tt STARLIGHT} appear to be in good agreement with those in \cite{Goddard-17} as well. In contrast to \cite{Goddard-17}, we have found rather weak gradients in stellar metallicity for massive SF and PQ galaxies. In a parallel MaNGA paper led by Hongyu Li et al. (in prep.), we have done detailed comparisons between {\tt STARLIGHT} and {\tt pPXF}, finding them to be consistent with each other in terms of both stellar age and metallicity gradients. The discrepancy with \cite{Goddard-17} is probably due to the different spectral fitting procedures adopted in their code; more work is needed to ascertain the source of the discrepancy. In any case, the large gradients of \dindex\ as shown for massive galaxies in our work is consistent with the predictions of most of the commonly-used models, and so are unlikely to be caused by stellar metallicity gradients.  

\cite{Kauffmann-15} analyzed the recent star formation history of low-mass galaxies from SDSS with stellar masses in the range $10^8-10^{10}$\msolar, using \dindex\ and \ewhda\ in combination with SFR$/$\mstar\ as diagnostics. It was suggested that a large fraction of the SFR density in these galaxies was contributed by multiple starbursts triggered by gas cooling and supernova feedback cycles over the history of the galaxy. The current work does not attempt to derive any model-dependent SFH parameters such as $F_{burst}$, the fraction of stellar mass formed in bursts. We have analyzed both \dindex\ and \ewhda, and the third parameter \ewhae\ is expected to be correlated with SFR/\mstar. Therefore, our measurements of the radial profiles for these diagnostics should in principle provide useful constraints on the picture proposed by \cite{Kauffmann-15}, which was limited by the single-fibre spectroscopy from SDSS, thus probing the stellar populations only within the central 1-2 kpc of a galaxy. Although the continuous star formation models of BC03 can explain the recent SFH for the majority of our galaxies in the same mass range, a model with multiple starbursts cannot be simply ruled out. It would be interesting to see whether their model and the continuous star formation models can be discriminated using the resolved spectroscopy from MaNGA in a future work.

\subsection{A critical mass for ``inside-out'' star formation cessation}

The radial gradients in the SFH diagnostics as measured from the MaNGA galaxies have revealed an ``inside-out'' picture of star formation cessation. This can be seen from both the radial profiles of the three diagnostic parameters as shown in Figure \ref{fig:D4000_profile}, and the comparisons of the inner and outer \dindex\ as shown in Figure~\ref{fig:D4000_in_out}. The diagnostic diagrams examining the mutual relations of the three parameters as shown in Figures~\ref{fig:diagram_one}-\ref{fig:diagram_three} further indicate that this inside-out cessation must happen in a rather smooth manner, with star formation rate declining continuously on a long timescale. In this picture, as discussed in \S\ref{sec:d4000_in_out}, a fully star-forming galaxy with a low stellar mass and weak/no gradients in \dindex\ will evolve into the partly quenched phase with a quenched inner region and a star-forming outer region, thus displaying significantly negative gradients in stellar age, and eventually to the final, fully quenched stage with a larger stellar mass and only weak/no gradients again. 

When we limit the analysis to a narrow range of stellar mass (see Figure~\ref{fig:D4000_in_out_mass}), we find the picture of inside-out cessation appears to hold only for galaxies with stellar masses above $\sim10^{10}$\msolar. For less massive galaxies, the inner and outer regions seem to evolve synchronously, showing almost no gradients at all in the evolutionary stages as classified by the $f_{Q}$ parameter. This finding implies that the inside-out cessation is at work in a galaxy only when its stellar mass exceeds a critical mass, which is a few $\times10^{10}$\msolar. A galaxy with an initial stellar mass below this critical mass may evolve to its fate via two different paths, depending on the final mass. If the final mass is below the critical mass, the star formation would cease synchronously across the galaxy, thus showing no/weak radial gradients over the galaxy lifetime. If the galaxy manages to keep forming stars over a long timescale so that its stellar mass exceeds the critical mass, or if the galaxy begins with a mass above the critical value, the inside-out star formation cessation would then be at work. For a fully quenched galaxy with current mass below the critical value, the star formation must have ceased in a synchronous manner with no radial dependence. 

We have further examined whether the onset of inside-out cessation depends on the bulge-to-total luminosity ratio and the spheroid/disk classification. It is interesting that, although the relative fraction of the galaxies falling in the subsets of SF, PQ and TQ may vary according to $f_{\rm B/T}$ or morphology, the overall distribution of galaxies on the \dindex$_{1.5Re}$-\dindex$_{\rm cen}$ plane appear to depend on stellar mass only. At higher masses, all types of galaxies show larger radial variations in \dindex, regardless of their morphology type or bulge-to-total ratio. This analysis demonstrates that stellar mass is indeed a driving parameter for the star formation cessation process in galaxies.

\subsection{Galactic bulge: reason or result of the star formation cessation?}

%\begin{figure*}
%\begin{center}
%\epsfig{figure=./fig/profile_mass_fqr15_new_wei.eps,clip=true,width=1.0\textwidth}
%\end{center}
%\caption{The median surface mass density profiles for SF (blue solid curves), PQ (green solid curves) and 
%TQ galaxies (red solid curves) in 5 stellar mass bins. The dashed curves represent the 20 and 80 per cent percentile 
%lines. We present the radial profiles with radii scaled the effective radius. }
%\label{fig:mass_prof}
%\end{figure*}

One might argue that the results regarding the inside-out cessation as discussed above can be interpreted by the increasing predominance of galactic bulges with increasing stellar mass, which may naturally explain the \dindex\ gradients observed in massive galaxies. Figure~\ref{fig:dependence_bulge} shows that, the galaxies with a small bulge luminosity fraction ($f_{\rm B/T}<0.4$) present stronger \dindex\ gradients at higher masses, in exactly the same way as seen for the bulge-dominated galaxies with $f_{\rm B/T}>0.4$. Therefore, the inside-out cessation of star formation as observed in the massive galaxies cannot be solely attributed to bulge growth. In the Appendix, we present an additional analysis in which we select galaxies with bulge radius $R_{\rm bulge}$ smaller than 0.5\Reff, thus estimating \dindex\ gradients using only the disk-dominated regions, and we find significant mass-dependent gradients similar to those seen in the previous section.

Previous studies have established that the presence of a central dense object (such as a prominent bulge) is a necessary, but not sufficient, condition for a central galaxy to be quenched \citep[e.g.][]{Bell-08, Bell-12, Fang-13}. \citet{Bluck-14} further found the bulge mass to play a more important role in quenching star formation in local galaxies when compared to the bulge-to-total stellar mass ratio. In agreement with these previous studies, Figure~\ref{fig:dependence_bulge} shows that the majority of the totally quenched galaxies in our sample have $f_{\rm B/T}>0.4$, but on the other hand only a fraction of the galaxies with $f_{\rm B/T}>0.4$ are fully quenched. 

However, we would like to point out that this agreement doesn't necessarily mean that a massive bulge is a necessary condition, or a driver for star formation quenching. Rather, these results show that the bulge is just a natural byproduct of the evolution process. This can be clearly seen from Figure~\ref{fig:dependence_bulge} where deviations of massive galaxies from the 1:1 relation, indicative of inside-out cessation, can happen in all the galaxies with central \dindex\ larger than $\sim1.4$, no matter whether the galaxy has a prominent bulge or not. In fact, the deviation from the 1:1 relation happens quite early when the galaxy is at the star-forming stage without a prominent bulge, and a large bulge-to-total ratio is observed mostly when it is approaching the final stage to become a fully quenched galaxy. If a prominent bulge is a necessary condition for driving the cessation process, the deviation should be observed only in the upper panels of that figure where the galaxies have a large $f_{\rm B/T}$. For less massive galaxies which may not follow the inside-out cessation process, a large $f_{\rm B/T}$ is unlikely a necessary condition, either, as we do see TQ galaxies in the lower panels with $f_{\rm B/T}<0.4$, which are not a negligible population considering the small number of PQ galaxies in both upper and lower panels at low masses. 

We conclude that the presence of a central dense object such as a bulge is a byproduct, but not a driving factor of the star formation cessation in local galaxies. It is possible that, when fully grown, the bulge can play some role in finalizing the cessation process, perhaps by suppressing bar formation or even destroying the bar structure in the galactic center which is believed to be the primary driver of the secular evolution. Numerical studies predicted that the central random motions can stabilize the disk against bar formation \citep{Athanassoula-Sellwood-86, Athanassoula-08}, and an anti-correlation of bar fraction with central stellar velocity dispersion has been observed for both low-z and high-z galaxies \citep{Das-08, Sheth-12, Cervantes-Sodi-13}.

\section{Conclusions}

We have investigated the resolved star formation histories of 1917 local galaxies with regular morphology using integral field spectroscopy from the MaNGA MPL5. We have obtained both two-dimensional maps and radial profiles of three parameters, \dindex, \ewhda\ and \ewhae, which combine to provide powerful diagnostics of the recent star formation history. Based on the \dindex\ and \ewhda\ maps,  we propose a new parameter, $f_Q(1.5R_e)$, to quantify the overall star formation status of a galaxy, defined as the stellar mass-weighted fraction of the quenched spaxels within $1.5R_e$. Accordingly we divide our galaxies into three subsets: fully star-forming (SF,  $f_Q(1.5R_e)<0.1$), partly quenched (PQ, $0.1\leq f_Q(1.5R_e)<0.9$) and totally quenched (TQ, $f_Q(1.5R_e)\geq 0.9$). We then study the radial gradients in the SFH diagnostic parameters, comparing the results for the three types and for galaxies with different stellar masses and structural parameters.

Our conclusions can be summarized as follows.

\begin{itemize}

\item Less massive galaxies with stellar mass below $\sim10^{10}$\msolar\ present no or very weak gradients in all the diagnostic parameters, regardless of the overall star formation status as quantified by $f_Q(1.5R_e)$, but the amplitudes of the profiles vary with $f_Q(1.5R_e)$, with older stellar populations being more dominant at larger $f_Q$. 

\item Massive galaxies with stellar mass above $\sim10^{10}$\msolar\ present significant gradients in all the three diagnostics if classified as an SF or PQ galaxy, but show weak gradients in both \dindex\ and \ewhda\ and no gradients in \ewhae\ if classified as a TQ galaxy. The observed gradients are stronger with increasing mass. 

\item The majority of the spaxels of the sample galaxies closely follow the continuous star formation models of \cite{Bruzual-Charlot-03}, regardless of stellar mass and $f_Q(1.5R_e)$, indicating that ongoing starbursts are very rare in galaxies with regular morphology. 

\item Distributions of the SF, PQ and TQ populations on the plane of central \dindex\ versus \dindex\ gradient, or the plane of central \dindex\ versus the \dindex\ at 1.5\Reff, reveal a critical stellar mass, $\sim10^{10}$\msolar, above which the star formation cessation in a galaxy happens from inside out. Galaxies tend to evolve synchronously at all radii, before their mass reaches the critical mass. 

\item The above conclusion holds for galaxies with or without a significant bulge, and for galaxies with spheroid- or disk-like morphologies. This indicates that the presence of a central dense object is not a driving parameter, but rather a byproduct of the star formation cessation process.

\end{itemize}

The large sample of galaxies with integral field spectroscopy from the ongoing MaNGA survey has enabled us to extensively analyze the radial gradients in the recent SFH of local galaxies and examine the dependence on both stellar mass and structural properties. This has led to the finding of a critical mass, which divides galaxies into two distinct classes following different star formation cessation processes. We would like to emphasize that more work is needed in order to better understand the implications and limitations of our results which are limited to galaxies with regular morphologies. That will be the purpose of the next papers in this series.

\acknowledgments
This work is supported by National Key Basic Research Program of China (No. 2015CB857004). EW is supported by the Youth Innovation Fund by University of Science and Technology of China (No. WK2030220019), the China Postdoctoral Science Foundation (No. BH2030000040) and NSFC (Grant No. 11421303). CL acknowledges the support of NSFC (Grant No. 11173045, 11233005, 11325314, 11320101002) and the Strategic Priority Research Program ``The Emergence of Cosmological Structures'' of CAS (Grant No. XDB09000000). MAB acknowledges support from NSF AST-1517006. RR thanks to CNPq and FAPERGS for partial financial support.

Funding for the Sloan Digital Sky Survey IV has been provided by the Alfred P. Sloan Foundation, the U.S. Department of Energy Office of Science, and the Participating Institutions. SDSS- IV acknowledges support and resources from the Center for High-Performance Computing at the University of Utah. The SDSS web site is www.sdss.org.

SDSS-IV is managed by the Astrophysical Research Consortium for the Participating Institutions of the SDSS Collaboration including the Brazilian Participation Group, the Carnegie Institution for Science, Carnegie Mellon University, the Chilean Participation Group, the French Participation Group, Harvard-Smithsonian Center for Astrophysics, Instituto de Astrofísica de Canarias, The Johns Hopkins University, Kavli Institute for the Physics and Mathematics of the Universe (IPMU) / University of Tokyo, Lawrence Berkeley National Laboratory, Leibniz Institut für Astrophysik Potsdam (AIP), Max-Planck-Institut für Astronomie (MPIA Heidelberg), Max-Planck-Institut für Astrophysik (MPA Garching), Max-Planck-Institut für Extraterrestrische Physik (MPE), National Astronomical Observatory of China, New Mexico State University, New York University, University of Notre Dame, Observatório Nacional / MCTI, The Ohio State University, Pennsylvania State University, Shanghai Astronomical Observatory, United Kingdom Participation Group, Universidad Nacional Autónoma de México, University of Arizona, University of Colorado Boulder, University of Oxford, University of Portsmouth, University of Utah, University of Virginia, University of Washington, University of Wisconsin, Vanderbilt University, and Yale University.

\bibliography{rewritebib.bib}

\appendix
%\section{Appendix}
\subsection{The metallicity and stellar age profiles from {\tt STARLIGHT}}
\begin{figure*}
 \begin{center}
    \epsfig{figure=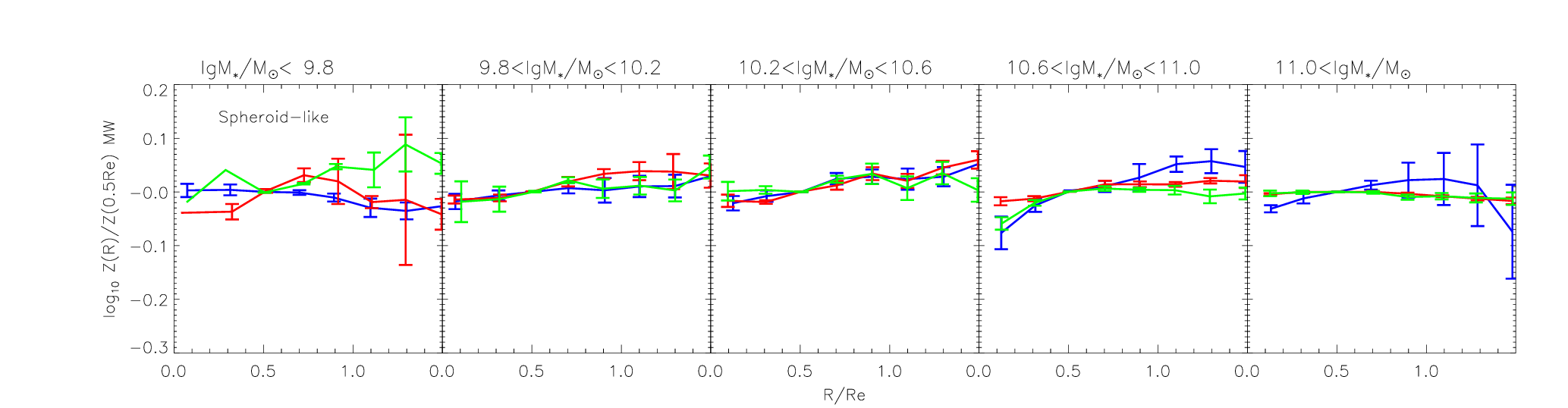,clip=true,width=1.0\textwidth}
    \epsfig{figure=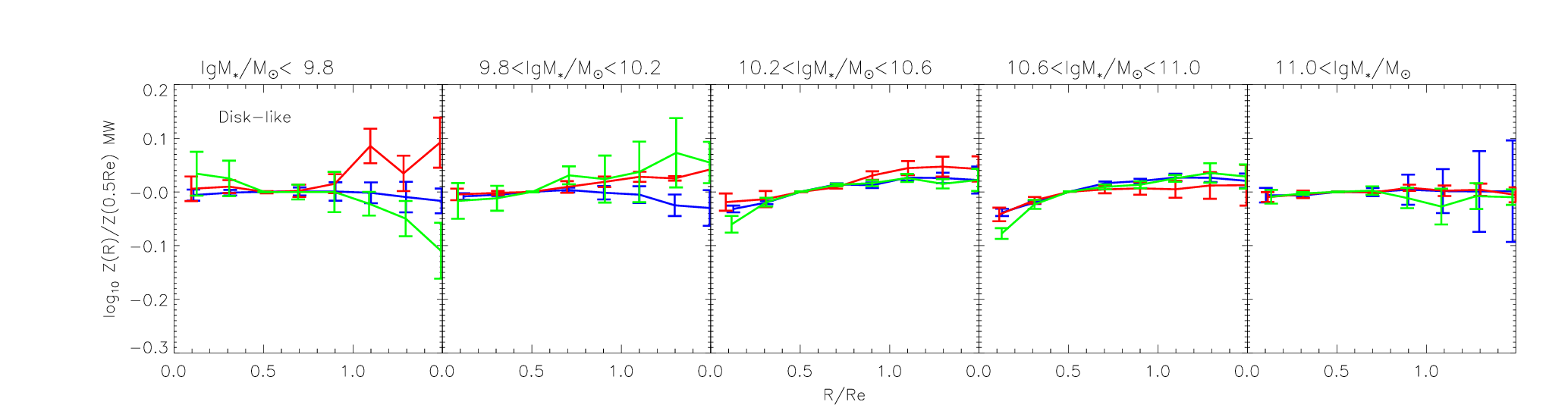,clip=true,width=1.0\textwidth}
   \end{center}
   \caption{The median radial gradients of mass-weighted stellar metallicity
    for SF (blue curves), PQ (green curves) and TQ (red curves) populations. The gradients
   are displayed with increasing stellar mass bins from left to right for spheroid-like (upper panels) and disk-like (bottom panels) galaxies. The colors and symbols are
   the same as Figure \ref{fig:three_grad}.  The errors are measured by bootstrap method.}
   \label{fig:metal_prof}
\end{figure*}

In Figure \ref{fig:metal_prof}, we present the gradients of mass-weighted stellar metallicity for SF (blue curves), PQ (green curves) and TQ (red curves) galaxies in the five stellar mass bins. The gradients of metallicity are presented for spheroid-like and disk-like galaxies separately. 
The stellar metallicities are from the {\tt STARLIGHT} output. As shown, no or weak
radial variations in stellar metallicity are found in almost all the subsamples. 
Both spheroid-like and disk-like galaxies with $10.2<$\lgmstar$<11.0$ are found to have slightly positive gradients regardless of their quenching statuses, while 
galaxies with \lgmstar$<$9.8 and \lgmstar$>$11.0 appear to have flat radial profiles of stellar metallicity as a whole.
We note that it is difficult to measure the stellar metallicity in an accurate way, since the measurements of stellar metallicity largely depend on the spectral fitting procedures. 

\begin{figure*}
  \begin{center}
    \epsfig{figure=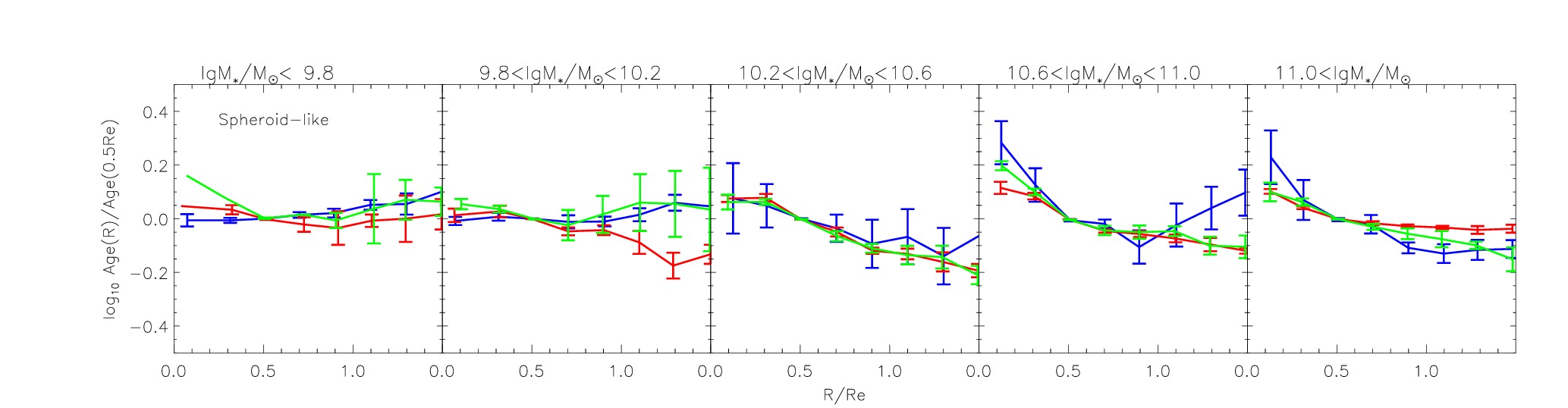,clip=true,width=1.0\textwidth}
    \epsfig{figure=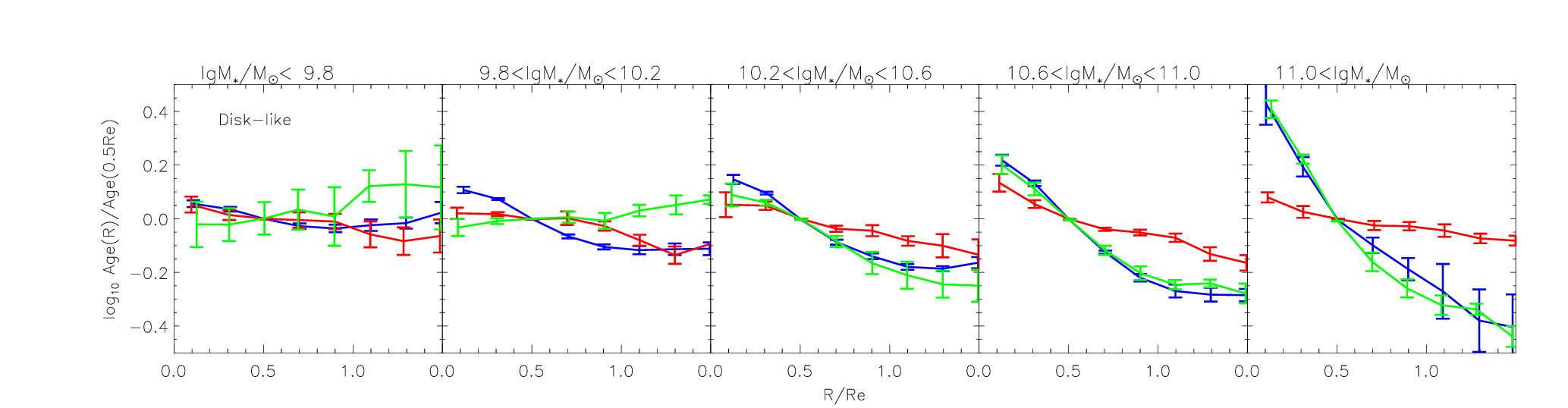,clip=true,width=1.0\textwidth}
  \end{center}
  \caption{The median radial gradients of light-weighted stellar ages for SF (blue curves), PQ (green curves)
   and TQ (red curves) populations. The gradients
   are displayed with increasing stellar mass bins from left to right for spheroid-like (upper panels) and disk-like (bottom panels) galaxies. The colors and symbols are the same as
   Figure \ref{fig:three_grad}.  The errors are measured by bootstrap method.}
   \label{fig:age_prof}
\end{figure*}

Since all our results are based on the three diagnostic parameters, here we try to examine our results by using the mean stellar age. 
Figure \ref{fig:age_prof} shows the median radial gradients of light-weighted stellar ages for SF (blue lines), 
PQ (green lines) and TQ (red lines) galaxies for the five stellar mass bins. Similarly, we present these for spheroid-like and disk-like galaxies separately. 
The light-weighted stellar ages are derived from {\tt STARLIGHT} measurement.  
As can be seen from Figure \ref{fig:age_prof}, 
the stellar age of disk-like TQ galaxies exhibits weak radial variations for both low and high stellar mass bins.
In contrast, the disk-like SF and PQ galaxies show pronounced age gradients for high stellar mass bins, 
while show no or weak radial variations for low stellar mass bins. 
In addition, these age gradients increase with increasing stellar mass monotonously. However, the spheroid-like galaxies exhibit weak radial variations of stellar age, and their stellar age gradients do not show significant differences for SF, PQ and TQ populations. 
In general, all these results are well consistent with what we have found 
in Figure \ref{fig:D4000_profile} and Figure \ref{fig:three_grad}. 

\begin{figure*}
  \begin{center}
    \epsfig{figure=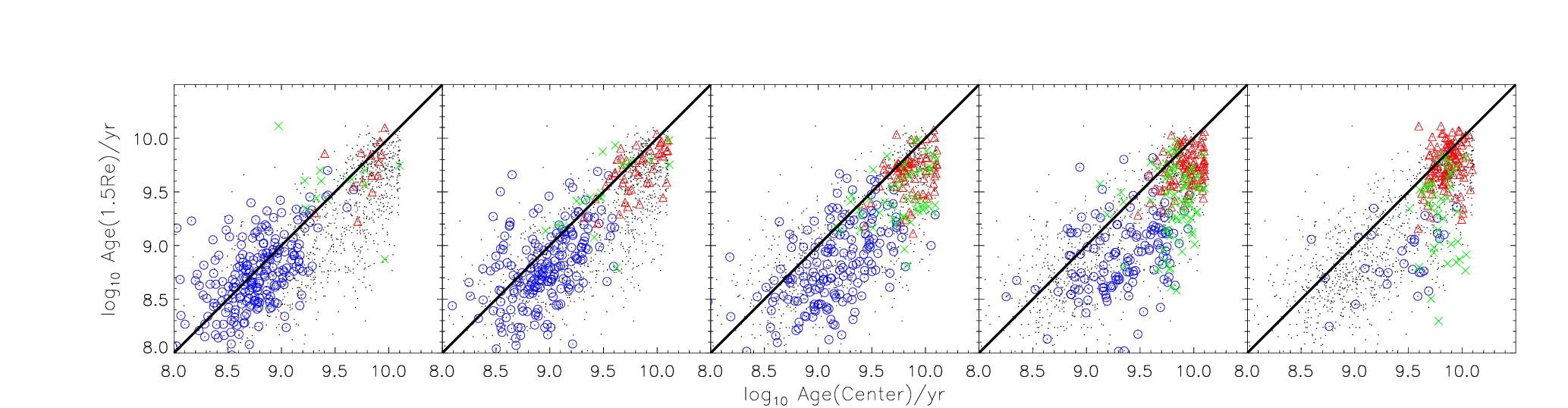, clip=true, width=1.0\textwidth}
    \epsfig{figure=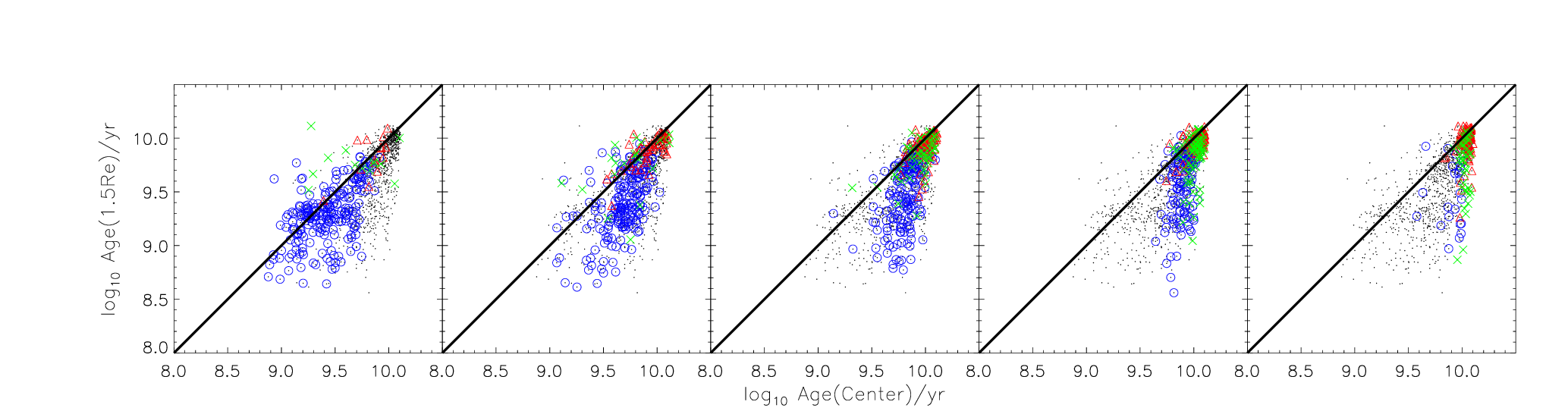,clip=true,width=1.0\textwidth}
  \end{center}
  \caption{The comparison between mean stellar age at the galactic center and 1.5\Reff\ radial bin 
   in 5 stellar mass bins.  We present these comparisons
   by using light-weighted (top panels) and mass-weighted (bottom panels) stellar age, respectively.
   Colors and symbols are the same as Figure \ref{fig:D4000_in_out_mass}. The black line in each panel is the 
   one-to-one line. For comparison, all galaxies in our sample are plotted as small gray dots in each panel.}
   \label{fig:age_in_out_mass}
 \end{figure*}

In addition, we compare the mean stellar age at galactic centers with at 1.5\Reff\ radial bins, 
shown in Figure \ref{fig:age_in_out_mass}. 
We present the light-weighted age comparisons in top row, and mass-weighted age 
comparisons in bottom row for SF, PQ and TQ galaxies in the same five stellar mass bins. 
As expected, the mean stellar age increases from SF, through PQ to TQ populations in both high and low
stellar mass bins.  Thus, we can illustrate these panels with assuming galaxy evolves from SF, through PQ 
to TQ status. As seen from the top row, the stellar populations in inner and outer regions 
become old simultaneously for less massive galaxies. However, for massive galaxies, 
the stellar population in central regions get old first, then extend to larger and larger radii.  
This confirms our previous result that star formation cessation occurs in inner and outer regions
 simultaneously for less massive galaxies, while it occurs firstly in galactic centers, then extend
to larger and larger radii for massive galaxies. 

In the bottom row of panels of Figure \ref{fig:age_in_out_mass}, there is a good linear relation between the 
mass-weighted age at galactic center
and 1.5\Reff\ radial bins in the logarithmic space in each stellar mass bin. 
With increasing stellar mass, this relation becomes steeper and steeper and the scatter gets to be smaller and smaller in the logarithmic space.
For SF galaxies, mass-weighted stellar ages are significantly larger in galactic center than that of 
1.5\Reff\ radial bins,  especially for massive galaxies ($\sim 10.0$ Gyr). 
However, this is not the same case for light-weighted stellar ages. 
Although some SF galaxies are forming new stars in their centers,
there has already been a dominated fraction of old stars formed in the past several Gyrs ago. 
Compared to their inner regions, the outer regions are dominated by young stars, which reveals an inside-out growth scenario.

We note that the flat metallicity profiles appear to be inconsistent with the previous results in the literature. However, we have verified that this issue is not caused by the MaNGA data itself. For this, we have reproduced Figures~\ref{fig:metal_prof}-\ref{fig:age_in_out_mass}, using the stellar age and metallicity maps produced by the Pipe3D pipeline \citep{Sanchez-16a,Sanchez-16b}, which are publicly available as a value-added catalog from the SDSS data release 13 at http://www.sdss.org/dr14/manga/manga-data/manga-pipe3d-value-added-catalog/. We find negative metallicity gradients for massive galaxies, consistent with previous results. More importantly, the stellar age gradients are in good agreement with those from STARLIGHT, indicating that the D4000 gradients presented in the main text are not dominated by possible gradients in metallicity.

 \subsection{The \dindex\ gradients for disk dominated regions}

  \begin{figure*}
  \begin{center}
    \epsfig{figure=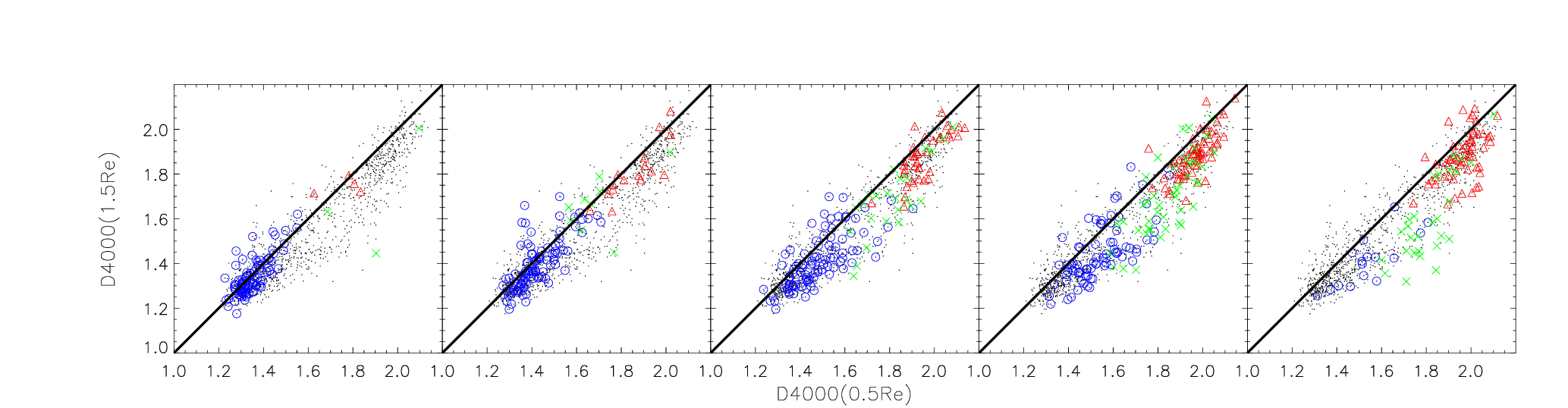,clip=true,width=1.0\textwidth}
  \end{center}
  \caption{The \dindex$_{\rm 0.5Re}$ vs. \dindex$_{\rm 1.5Re}$ plane for galaxies with 0.5\re\ greater
   than $R_{\rm bulge}$.
  We present this figure with separating galaxies into the five stellar mass bins as above. The symbols, colors 
  and lines are the same as those in Figure \ref{fig:dependence_bulge}.}
   \label{fig:disk_in_out}
 \end{figure*}
 
As proposed in Section 4.3, quenched galaxies are usually found to have massive bulges.  One question arises:
whether the \dindex\ or stellar age gradients in massive SF and PQ galaxies are mainly due to the combination
 effect of a old massive bulge and a young stellar disk? To answer this question, we first select galaxies
  with 0.5\re\ greater than the bulge size ($R_{\rm bulge}$). The bulge size is taken from \cite{Simard-11}. 
 Then we compare the \dindex\ in 0.5\re\ and in 1.5\re\ radial bins to figure out whether the
  gradient in \dindex\ retains or not between 0.5\re\ and 1.5\re\ (the bin width is 0.2\re).  For the galaxies we selected, the 
  regions between 0.5\re\ and 1.5\re\ are dominated by disk component. 
Figure \ref{fig:disk_in_out} shows the \dindex$_{\rm 0.5Re}$ vs. \dindex$_{\rm 1.5Re}$ plane for galaxies
with 0.5\re\ greater than $R_{\rm bulge}$. We find that pronounced gradients still exist for massive SF and PQ galaxies 
even for disk dominated regions, which suggests that the existence of massive bulge is not the only reason
for the  observed \dindex\ gradients, and disk components also exhibit significant gradients in \dindex\ 
(or in light-weighted stellar age).    

\subsection{The definition of the quenched fraction  weighting by r-band flux}
\begin{figure*}
  \begin{center}
    \epsfig{figure=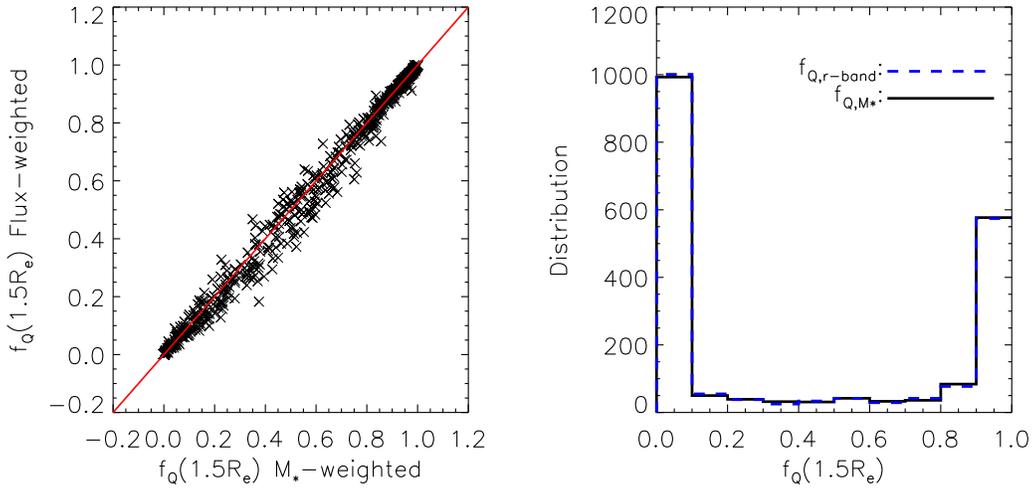,clip=true,width=0.8\textwidth}
  \end{center}
  \caption{Left panel: The comparison of $f_{\rm Q,M_*}$ vs $f_{\rm Q,flux}$. The red line is the one-to-one relation. Right panel: The distribution of $f_{\rm Q,M_*}(1.5Re)$ and $f_{\rm Q,flux}(1.5Re)$ for the sample galaxies. }
   \label{fig:fq_r_band}
 \end{figure*}

In the main text, we have introduced a parameter to describe the quenching status for each galaxy, the quenched fraction ($f_{\rm Q,M_*}$),  which is defined as the mass-weighted fraction of spaxels that are quenched within a given galactic radius. Since the stellar masses of individual spaxels are model dependent to some degree, we try to weight the spaxels with $r$-band flux in defining the quenched fraction ($f_{\rm Q,flux}$). The left panel of Figure \ref{fig:fq_r_band} shows the comparison of $f_{\rm Q,M_*}(1.5Re)$ and $f_{\rm Q,flux}(1.5Re)$ for the sample galaxies. As shown, the flux-weighted quenched fraction are in good linear correlation with the mass-weighted quenched fraction. We note that there are many data points in the low-$f_Q$ end and high-$f_Q$ end, since majority of galaxies are either SF or TQ ones. The right panel of Figure \ref{fig:fq_r_band} shows the distribution of $f_{\rm Q,M_*}(1.5Re)$ and $f_{\rm Q,flux}(1.5Re)$ for the sample galaxies. As shown, the distributions of the two are almost the same. This indicates that using the flux-weighted quenching definition would not change our results. In addition, we present a model independent definition of the quenched fraction weighting by $r$-band flux, which will be adopted in our following work.  

%%%%%%%%%%%%The End%%%%%%%%%%%%%%%%%%%%%%%%%%%%%%%%%%%%%%%%%
\label{lastpage}
\end{document}